\UseRawInputEncoding
\documentclass[useAMS,usenatbib]{mn2e}
\usepackage{epsfig}
\usepackage{graphicx}
\usepackage{amssymb}
\usepackage{color}
\usepackage{placeins}
\usepackage{subfigure}
\usepackage[fleqn]{amsmath}
\usepackage{threeparttable}
\usepackage{longtable}
\usepackage{booktabs}
\usepackage{caption}
\usepackage{float}
\usepackage[a4paper=true,dvipdfm=true,pagebackref=true]{hyperref}
\hypersetup{colorlinks = true, linkcolor = green, anchorcolor = red, citecolor = blue, filecolor = red, pagecolor = red, urlcolor = blue}
\usepackage{lineno}

\begin{document}
\title[The fundamental plane of blazars based on the black hole spin-mass energy]
  {The fundamental plane of blazars based on the black hole spin-mass energy}

\author[Zhang et al.]
{Xu Zhang$^{1}\thanks{zx@yxnu.edu.cn}$, Dingrong Xiong$^{2}\thanks{xiongdingrong@ynao.ac.cn}$, Quangui Gao$^{1, 3}$, Guiqin Yang$^{4}$\\
     \newauthor Fangwu Lu$^{1, 3}$, Weiwei Na$^{1}$, Longhua Qin$^{1}$\\
     $^1$ Department of Physics, Yuxi Normal University, Yuxi 653100, People's Republic of China\\
     $^2$ Yunnan Observatories, Chinese Academy of Sciences, 396 Yangfangwang, Guandu District, Kunming, 650216, People’s Republic of China\\
     $^3$ Key Laboratory of Astroparticle Physics of Yunnan Province, Kunming 650091, People's Republic of China \\
     $^4$ Faculty of Mechanical and Electrical Engineering, Kunming University of Science and Technology,\\ Kunming 650031, People's Republic of China}
\pagerange{\pageref{firstpage}--\pageref{lastpage}} \pubyear{2023}

\maketitle

\label{firstpage}

\begin{abstract}

We examine the fundamental plane of 91 Blazars which include FSRQs and BL Lacs with known X-ray luminosity ($L_{R}$), radio luminosity ($L_X$), and black hole mass measurements ($M$) to reflect the relationship between jet and accretion for blazars. The fundamental plane of Blazars are log$L_{R}$=${0.273}_{+0.059}^{-0.059}$log$L_X$+${0.695}_{+0.191}^{-0.191}$log$M$+${25.457}_{+2.728}^{-2.728}$ and log$L_{R}$=${0.190}_{+0.049}^{-0.049}$log$L_X$+${0.475}_{+0.157}^{-0.157}$log$M$+${28.568}_{+2.245}^{-2.245}$  after considering the effect of beam factor. Our results suggest that the jet of blazars has connection with accretion. We set the black hole spin energy as a new variable to correct the black hole mass and explore the effect of black hole spin on the fundamental relationship. We find that the fundamental plane of Blazars is effected by the black hole spin, which is similar to the previous work for AGNs. We additionally examine a new fundamental plane which is based on the black hole spin-mass energy ($M_{spin}$). The new fundamental plane (log$L_{R}$=${0.332}_{+0.081}^{-0.081}$log$L_X$+${0.502}_{+0.091}^{-0.091}$log$M_{spin}$+${22.606}_{+3.346}^{-3.346}$ with R-Square=0.575) shows that $M_{spin}$ has a better correlation coefficient comparing to the $M$ for fundamental plane of Blazars. These results support that the black hole spin should be considered as a important factor for the study of fundamental plane for Blazars. And these may further our understanding of the Blandford-Znajek process in blazars.

\end{abstract}



\section{INTRODUCTION}
When studying the internal structure of binary star systems and galactic cores, the existence of a black hole (BH) can be inferred through various indirect indicators. These indicators show clear associations with black hole activity. One such indicator is the synchrotron radiation emitted by relativistic jets in the radio frequency range, while another common indicator is the intense and compact power-law X-ray emission, often associated with the inner regions of accretion disks \citep{Begelman1984}. It is widely observed that the accretion of compact objects and the initiation of relativistic outflows/jets are interconnected phenomena \citep{Falcke1994}. This suggests an inherent correlation between the jet and disk accretion flux to some extent. To discuss this correlation, the concept of a fundamental plane relationship has been introduced. The fundamental plane for black hole activity is commonly defined as an approximate power-law relationship among the mass of BH, X-ray and radio luminosities. It usually applies to the hard state BHs. The applicability of fundamental plane for inclusion of soft or soft-intermediate state BHXBs, as well as Seyferts is still not sure. This relationship applies to black holes that produce jets \citep{Merloni2003,Falcke2004,Kording2006a,Kording2006b,Gultekin2009,Unal2020}. In black holes with outflows and an accretion disk, X-ray luminosity is generally linked to accretion power, while radio luminosity reflects the jet power in radiatively inefficient active galactic nuclei (AGNs) which have sub-Eddington accretion rates \citep{Heinz2003,Falcke2004,Merloni2003,Saikia2015,Kording2006a,Plotkin2012,Gultekin2009,Unal2020}.

Blazars are the most extreme type of active galactic nuclei (AGNs), with their jets pointing towards the observer \citep{Urry1995} and they constitute the brightest and most dominant population of AGNs in the gamma-ray band \citep{Fichtel1994,Abdo2010}. Their extreme observational properties are explained as a result of a beaming effect. Due to the relativistic beaming effect, the emission in the jet is strongly enhanced in the observer's frame \citep{Urry1995}. Blazars are typically divided into two sub-classes: BL Lacertae objects (BL Lacs) and flat spectrum radio quasars (FSRQs). FSRQs have strong emission lines, while BL Lacs objects have very weak or no emission lines. The classical division between FSRQs and BL Lacs is mainly based on the equivalent width (EW) of the emission lines. In a rest frame, objects with EW $>$ 5$\dot{A}$ are classified as FSRQs \citep[e.g.,][]{Urry1995,Scarpa1997}. \cite{Blandford1978} originally proposed that the absence of broad lines in BL Lacs is due to very bright, Doppler-enhanced continuous synchrotron radiation.

There is no consensus on the formation mechanism of relativistic jets until now. Three main mechanisms have been proposed to explain their formation. The first one is the Blandford-Znajek (BZ) mechanism, which posits that the rotational energy of the black hole and accretion disc are the primary source of the jets \citep{Blandford1977}. Furthermore, the jet power in the BZ mechanism is dependent on the spin of the black hole. The second mechanism is the Blandford-Payne (BP) mechanism, which suggests that the jet extracts the rotational energy of the accretion disc, and the mass of the black hole can be ignored. This mechanism has been supported by previous studies \citep{Rawlings1991,Ghisellini2010, Gu2009, Cao1999, Tavecchio2010,Sbarrato2012,Ghisellini2014,Xiong2015,Begelman1984}. Finally, A hybrid model for mixing BZ and BP mechanisms has been proposed by \cite{Meier2001} and \cite{Garofalo2010}, which could provide an explanation for the observed differences in AGNs with relativistic jets \citep{Garofalo2010}. The jet power of Blazars often exceeds the luminosity of the accretion disk \cite{Ghisellini2014}. It is shown that besides accretion, there are other important physical parameters that help to determine the jet power \cite{Meier2002}, such as the spin and black hole mass of supermassive black holes. In particular, the spin of a black hole has often been speculated as the main parameter for determining the jet power of BH \citep{Blandford1977}. \cite{Unal2020} have examined the fundamental plane which confirmed that the jet energy scales quadratically with BH spin of AGN. Considering the involvement of BH spin in both accretion and jets, BH spin is crucial for the study of fundamental plane\citep{Unal2020}.

As a special type of AGN, Blazars should theoretically conform to the fundamental plane relation, which has been confirmed by previous studies \citep{Merloni2003,Falcke2004,Bariuan2022,Unal2020,Saikia2015,Zhang2018}. \cite{Merloni2003} and \cite{Falcke2004} examined the fundamental plane with a large sample including the blazars and confirmed that Blazars are also follow the fundamental plane of AGN. \cite{Saikia2016} used the intrinsic radio luminosity obtained through the Lorentz factor to calculate the fundamental plane of Blazars. \cite{Zhang2018} selected the radiatively efficient sources to get the fundamental plane of Blazars. In this paper, we conduct a fundamental analysis of blazars, and compare them with normal radio-loud and radio-quiet AGNs \citep{Unal2020}. Furthermore, we use a new calculation method \citep{Daly2022} to incorporate black hole spin energy as an new variable to study the impact of black hole spin on the fundamental plane relation. The structure of this article is as follows. The sample is described in Section 2, and the analysis results are described in Section 3. Section 4 summarizes our conclusions.

\section{THE SAMPLES}
\begin{figure*}
  \centering
  \includegraphics[width=0.42\textwidth]{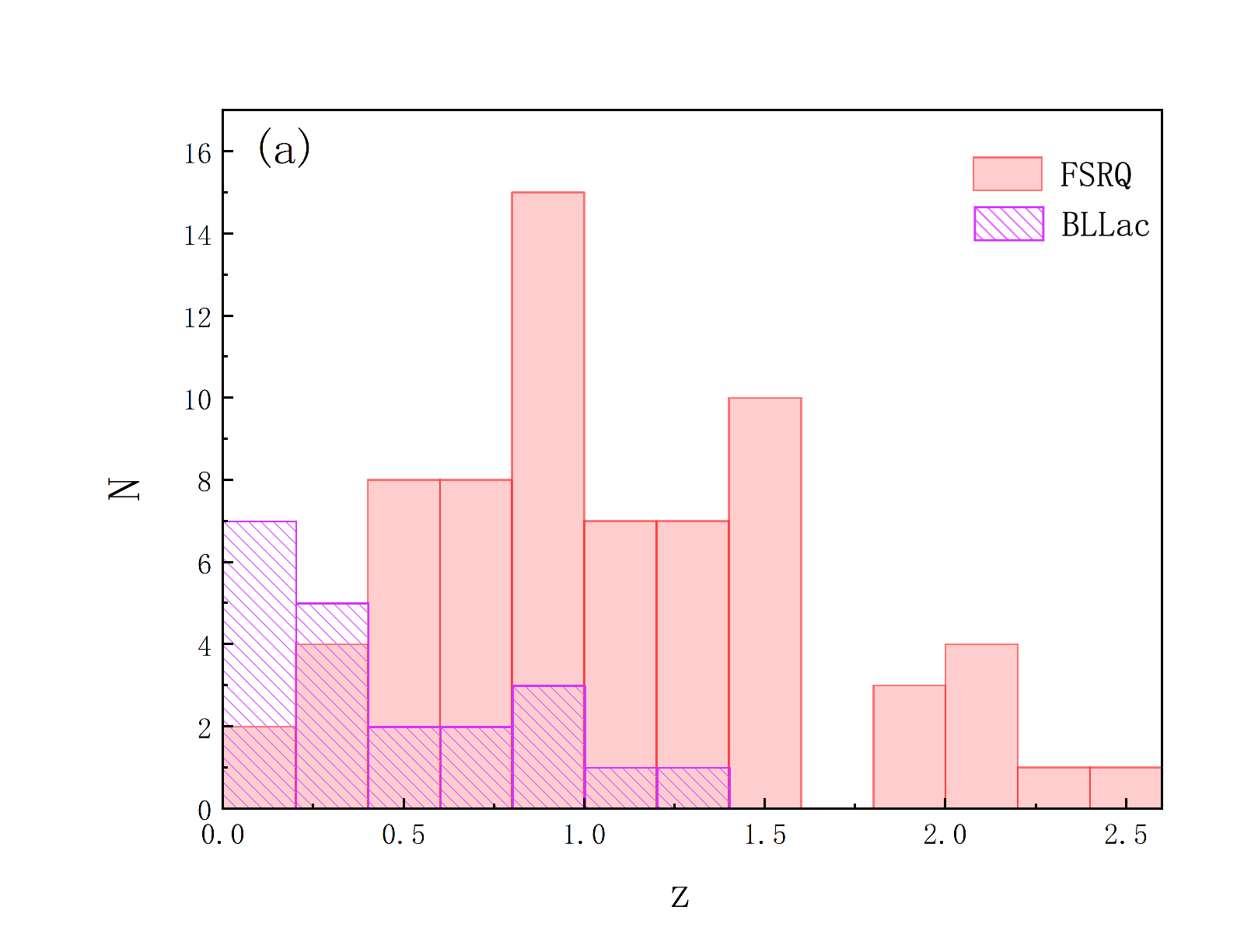}
  \includegraphics[width=0.42\textwidth]{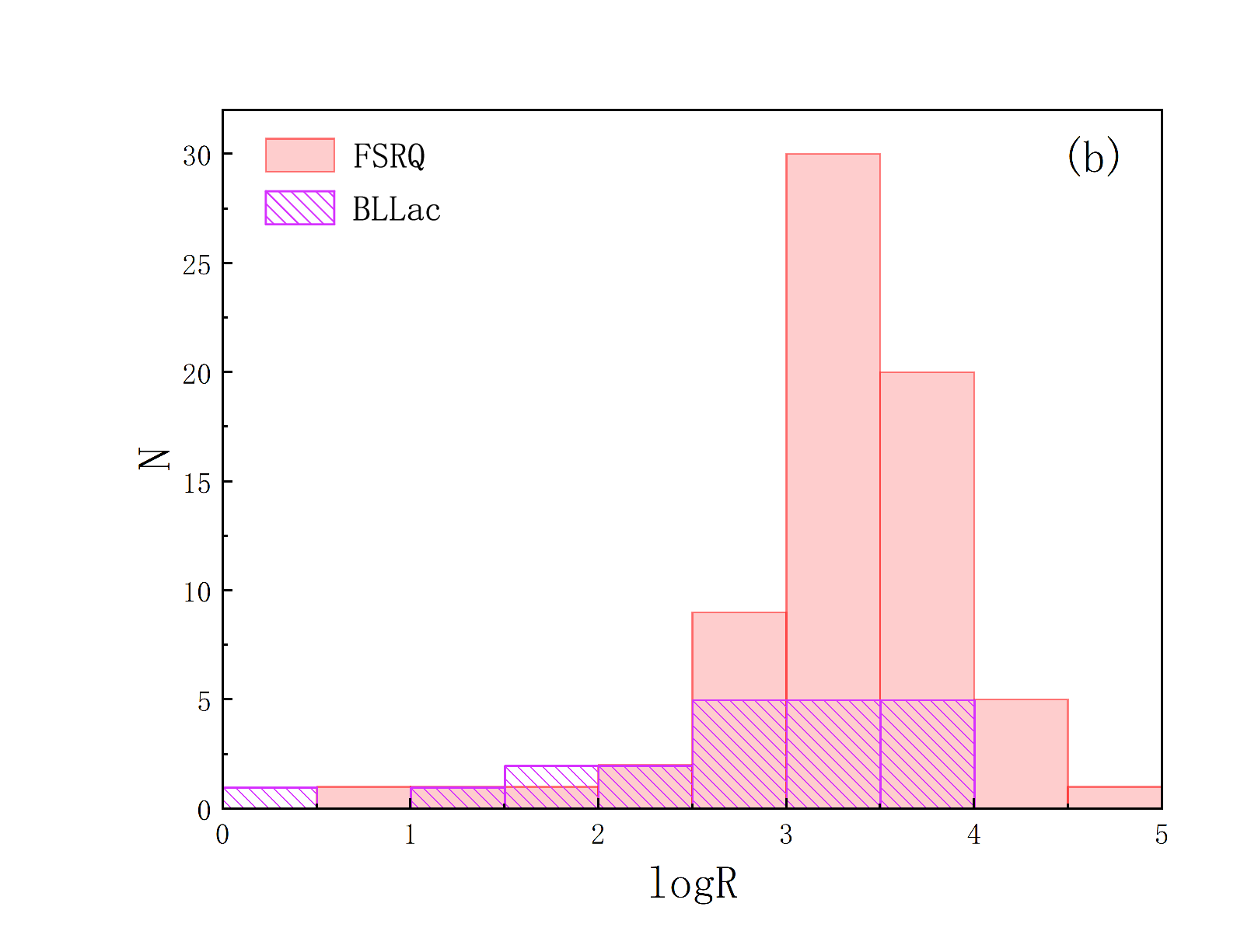}
  \includegraphics[width=0.42\textwidth]{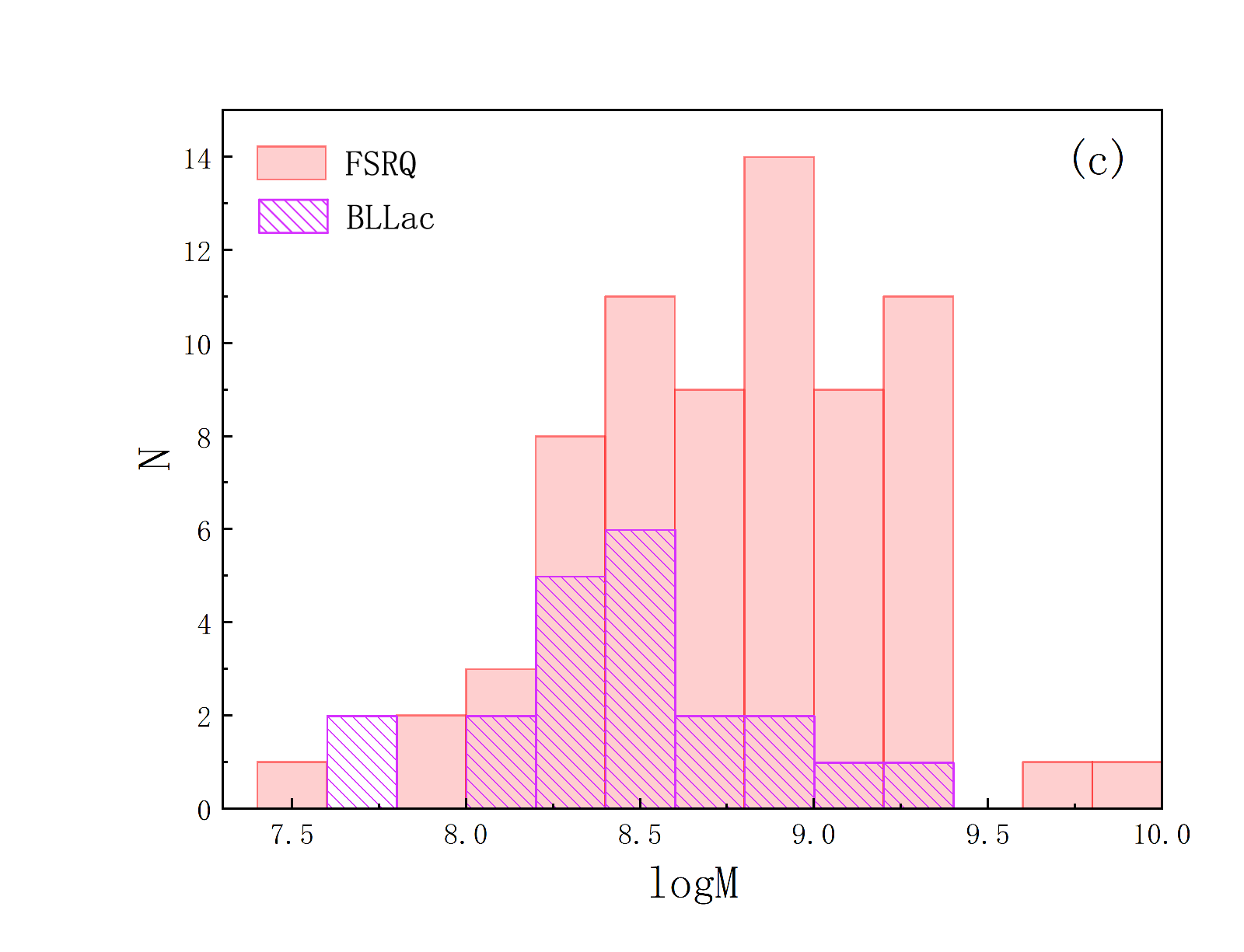}
  \includegraphics[width=0.42\textwidth]{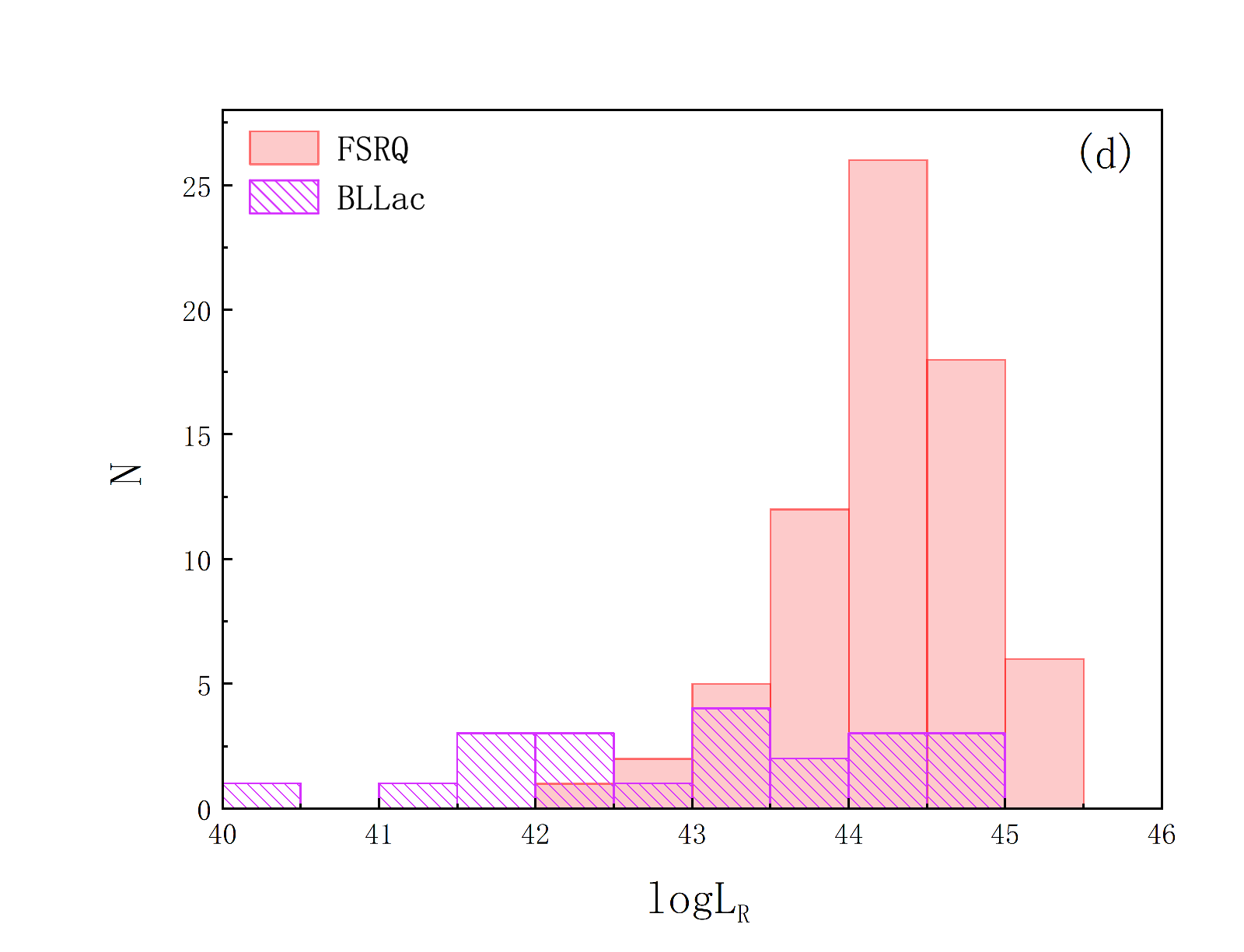}
  \includegraphics[width=0.42\textwidth]{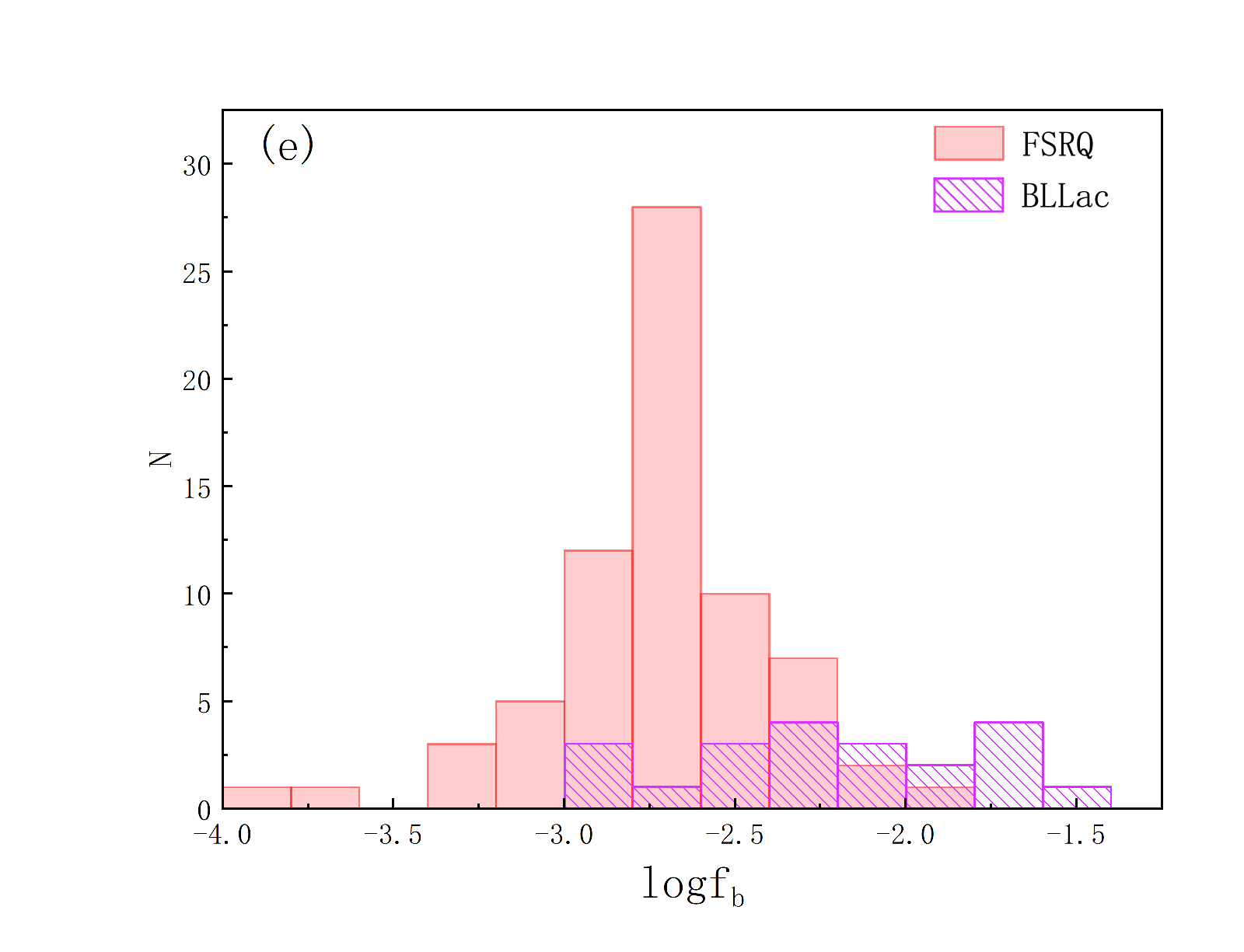}
  \includegraphics[width=0.42\textwidth]{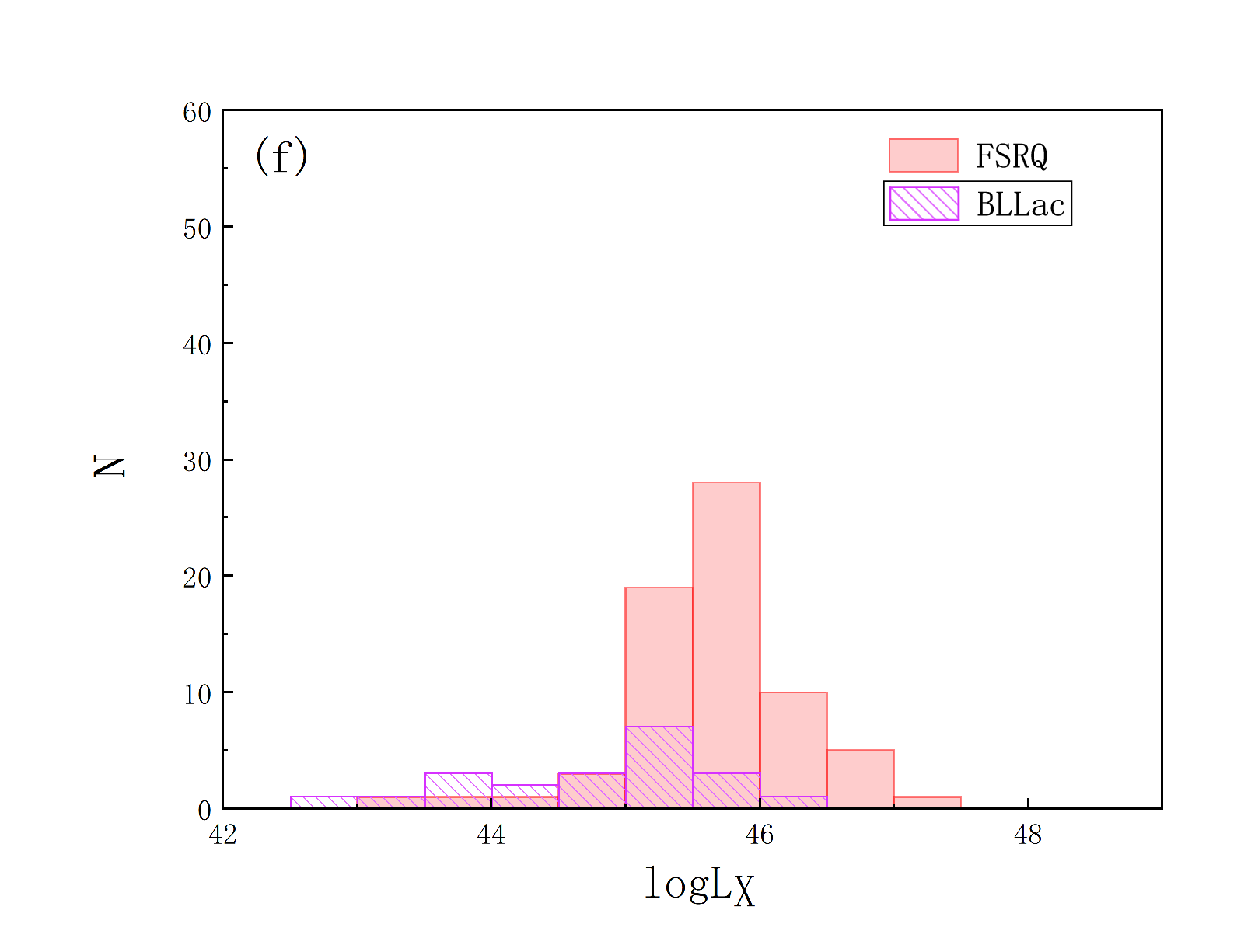}
  \includegraphics[width=0.42\textwidth]{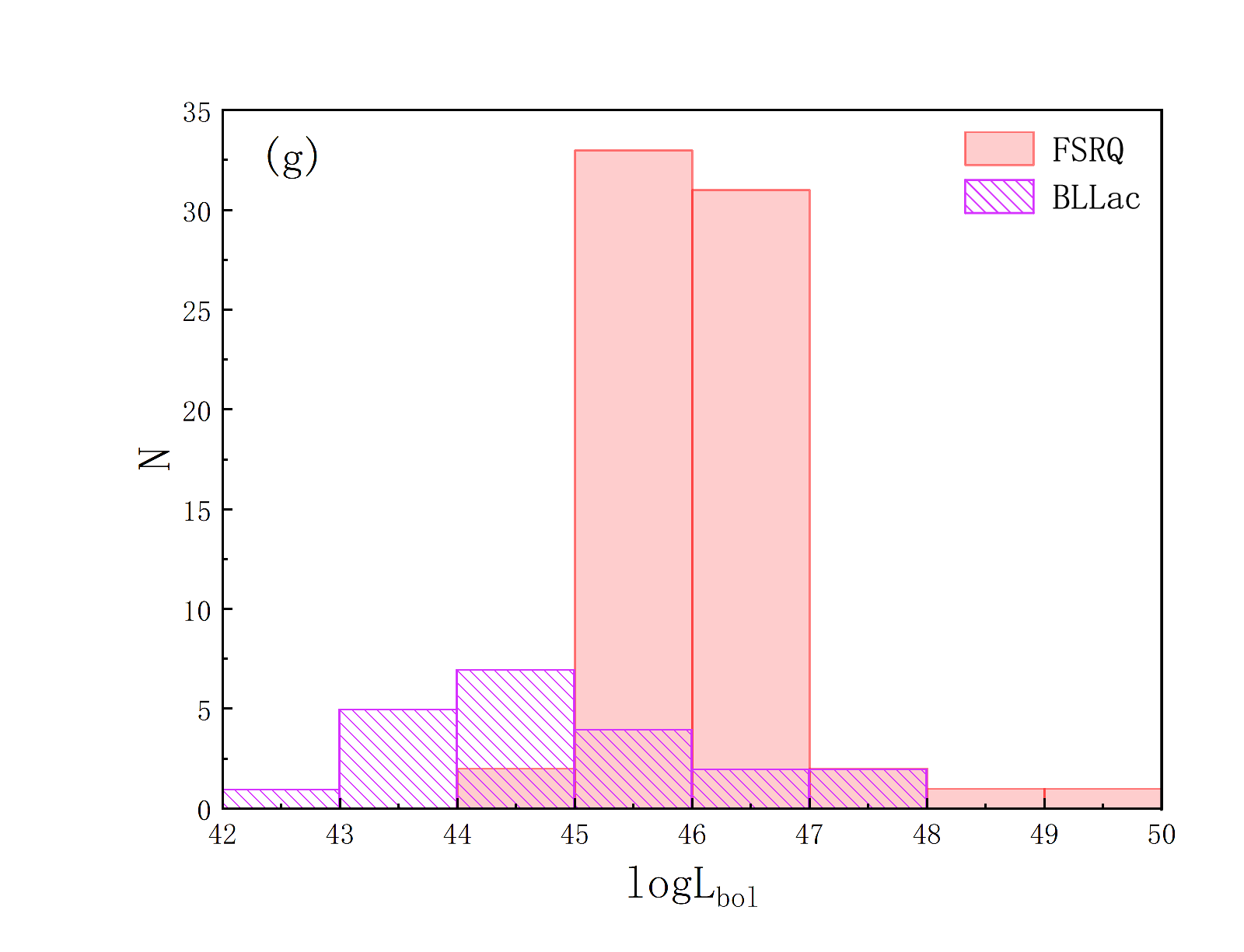}
  \includegraphics[width=0.42\textwidth]{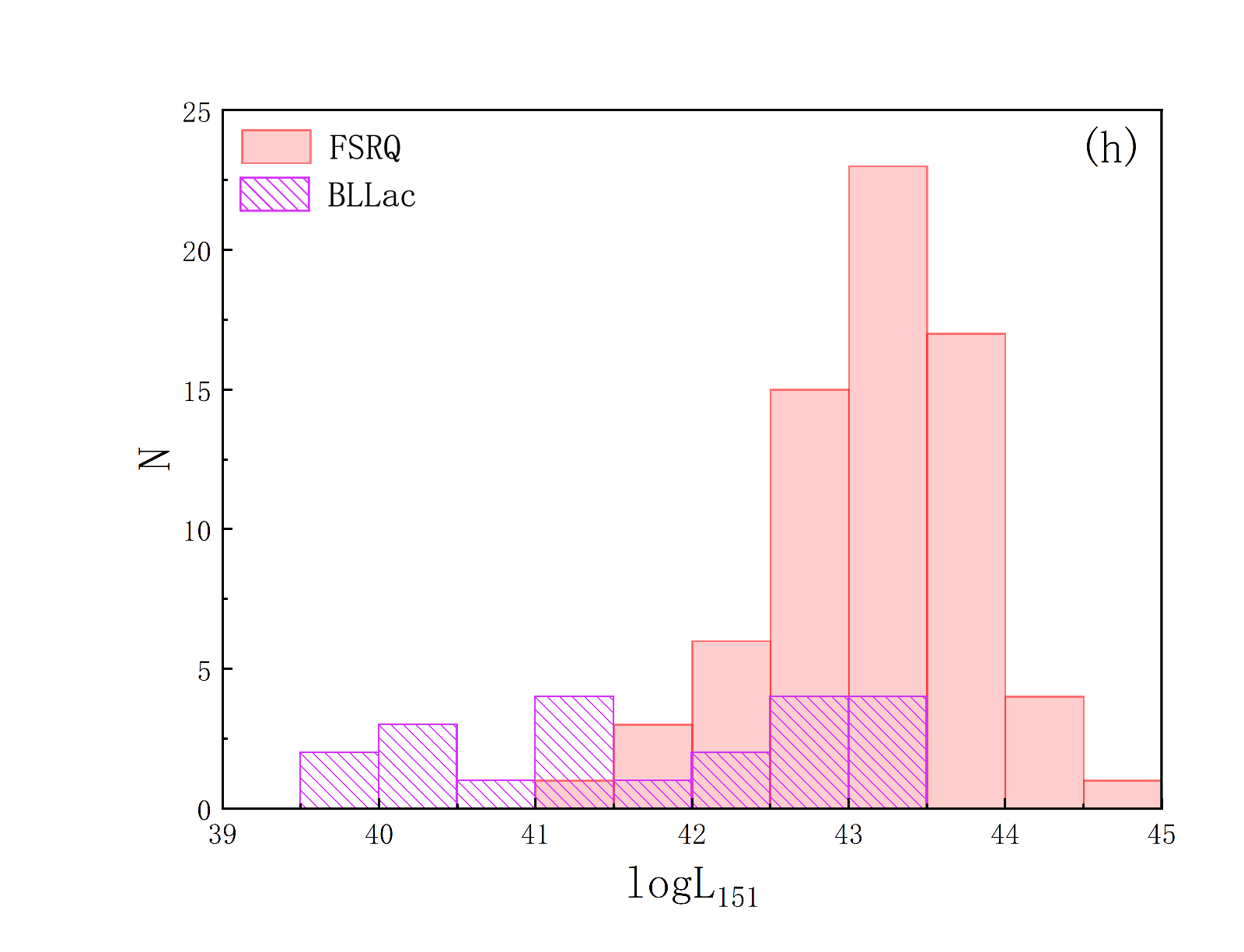}
  \caption{Distribution of the FSRQ and BL Lacs samples used in our study of fundamental plane for blazars. Histograms illustrate the distribution of several different physical properties, including:(a) redshift z, (b) radio loudness logR, (c) black hole mass log$M$, (d) 5 GHz radio luminosity log${L_{R}}$, (e) beam factor log$f_b$, (f) 2-10 keV X-ray luminosity log${L_{X}}$, (g) bolometric luminosity log${L_{bol}}$, (h) 151 MHz radio luminosity log${L_{151}}$. }
\label{fig:1}
\end{figure*}

In order to do the fundamental plane of blazars, we tried to collect the sample of blazars as much as possible with reliable redshift, 2-10 keV X-ray luminosity, 5 GHz radio luminosity,  BH mass, bolometric luminosity, beam power. For this purpose, we select the large samples of blazars from \cite{Ackermann2011} and \cite{Nolan2012} which can be detected by Very Large Array (VLA) for radio flux density and Swift or XMM-Newton for X-ray flux density. In order to collect as many sources as we need, we first obtain the samples following broad-line data of blazars \citep{Cao1999,Wang2004,Liu2006,Sbarrato2012,Chai2012,Shen2011,Shaw2012}. Secondly, we cross-correlated these sample with previous works \citep{Woo2002,Xie2004,Liu2006,Zhou2009,Zhang2012,Sbarrato2012,Chai2012,Leon2011a,Shen2011,Shaw2012} to obtain the BH mass. At last, we selected the sample of blazars with radio flux density and X-ray flux density which are detected by VLA and Swift or XMM-Newton on NASA/IPAC Extragalactic Database (NED). The beaming factor of sample in this paper can be collected by cross-matching  with the sources of \cite{Nemmen2012}. In total, we collected a sample containing 91 blazars which includes 70 FSRQs and 21 BL Lacs. All sample of 91 Blazars with redshift (z), ${\rm log}M_{\rm dyn}$, ${\rm log}L_{\rm X}$, ${\rm log}L_{\rm R}$, ${\rm log}L_{\rm {bol}}$, ${\rm log}f_{\rm b}$ and radio loudness (R) are listed in Table \ref{table:1}. We use the collected data to calculate the black hole spin energy ($M_{\rm spin}$) and irreducible black hole mass ($M_{\rm irr}$) of sample for further fundamental analysis.  All sample of Blazars with ${\rm log}L_{\rm j}$, ${\rm log}L_{\rm {EDD}}$, ${\rm log}{\rm F}$, ${\rm log}M_{\rm irr}$, ${\rm log}M_{\rm spin}$ are listed in Table \ref{table:2}. The sources are arranged in order of increasing redshift. The source distribution of Blazars' sample across several different physical properties is shown in Figure \ref{fig:1}. The z and ${\rm log R}$ distributions of the various classes are shown in part a and b of Figure \ref{fig:1}. The redshift distributions for all blazars are 0 $\leq z \leq$ 2.6 and mean value is 0.93. Mean values for FSRQs and BL Lacs are 1.07 and 0.44. The R distributions for all blazars are $10^{0.06}$ $\leq R \leq$ $10^{4.96}$ and mean value is $10^{3.16}$. Mean values for FSRQs and BL Lacs are $10^{3.16}$ and $10^{2.70}$.

\subsection{Black hole mass}
 The traditional virial BH mass also known as BH dynamical mass ($\rm{M_{dyn}}$) of AGN usually could be measured by the empirical relationship between the BLR size and the ionization luminosity, as well as the measured wide line width, assuming that the BLR cloud is gravitationally bound by the central black hole \citep{Xiong2014}. In this paper we collected the BH mass of FSRQs which are estimated by traditional virial method \citep{Woo2002,Wang2004,Liu2006,Shen2011, Chai2012,Sbarrato2012,Shaw2012}. If the BH mass of sample can be attained from different lines, we take the average of BH masses from different lines. The BH masses we got from \cite{Vestergaard2006} attained by H $\beta$ and $\rm{C_{IV}}$ lines, and \cite{Vestergaard2009} attained by Mg II lines. And for the BL Lac objects of sample, the masses of black holes can be estimated by the $M_{BH}$-$\sigma$ or $M_{BH}$-L relationship of their host galaxies, where $\sigma$ and L are the stellar velocity dispersion and the bulge luminosity of the host galaxy \citep{Xiong2014,Woo2002,Zhou2009,Leon2011a, Chai2012,Sbarrato2012,Zhang2012}. There is a little number of sources’ BH mass which are estimated by variation time-scale from \cite{Xie1991,Xie2004}. If there are more than one BH masses we got that belong to the same sample, then we take the average of those data. The average difference for BH mass between the method of traditional virial and the properties of host galaxies from the same source is 0.023. Due to the small number of black hole masses estimated by varying time-scales, the difference between the black hole masses obtained from \cite{Xie1991,Xie2004} and other methods has little effect on the results. The BH mass distributions of the various classes are shown in part c of Figure \ref{fig:1}. The BH mass distributions for all blazars are $10^{7.5}$ $\sim$ $10^{10}$ $\rm{M_{\odot}}$ and mean value is $10^{8.70\pm0.30}$ $\rm{M_{\odot}}$. Mean values for FSRQs and BL Lacs are $10^{8.78\pm0.30}$ $\rm{M_{\odot}}$  and  $10^{8.43\pm0.30}$ $\rm{M_{\odot}}$ .

\subsection{Intrinsic 5 GHz radio luminosity}
In this paper, 5 GHz radio luminosity was calculated by 5 GHz radio flux density from NED. These radio fluxes were taken from 5 GHz high resolution VLA observations directly. In order to ensure the accuracy of the collected data and reduce errors caused by converting through other approximate frequency, we only collect sources that have accurate 5 GHz radio flux density. At the same time, we strive to make the observation time of the collected 5 GHz radio data align as closely as possible with the observation time of other bands, reducing errors caused by variations in observation time. If there is more than one observation from the same source, we take the average value of all observations. For different observations with the same wavelength from the same source, the difference is usually controlled between 0.03 and 0.12. After we averaged the difference, the error was basically controlled within 0.05. Blazars usually have very powerful relativistic jets which includes gamma-ray band and radio band. For this reason, We should consider the influence of the beaming effect \citep{Xiong2014,Zhang2023a}. The intrinsic luminosity of radio luminosity can be estimated by beaming factor $f_{\rm{b}}$, {$f_{\rm{b}}$}$\equiv$1-cos{$\theta$} where $\theta$ is the jet opening angle \citep{Nemmen2012,Zhang2017}. The uncertainty of $f_{\rm{b}}$ we took is 0.15 as reported in \cite{Nemmen2012}.
\begin{equation}
\emph{$f_{\rm{b}}$}=1-\cos{\theta_{\rm{j}}},
\end{equation}
$\theta_{\rm{j}}$ is the jet opening angle of Blazars. The beaming factor $f_{\rm{b}}$ in this paper are obtained from \cite{Nemmen2012} and \cite{Xiong2014}. we can measure the intrinsic 5 GHz radio luminosity with beaming factor.
\begin{equation}
\begin{split}
L_{\rm{R^*}}=L_{\rm{R}}\emph{$f_{\rm{b}}$},
\end{split}
\end{equation}
$L_{\rm{R^*}}$ are the intrinsic luminosity of radio luminosity, respectively. The 5 GHz radio luminosity and beam factor distributions of the various classes are shown in part d and e of Figure \ref{fig:1}. The 5 GHz radio luminosity distributions for all blazars are $10^{40}$ $\sim$ $10^{46}$ erg $\rm{s^{-1}}$ and mean value is $10^{43.95\pm0.05}$ erg $\rm{s^{-1}}$. Mean values for FSRQs and BL Lacs are $10^{44.24\pm0.05}$ erg $\rm{s^{-1}}$ and $10^{42.99\pm0.05}$ erg $\rm{s^{-1}}$. The beam factor distributions for all blazars are -4 $\sim$ -1.25 and mean value is -2.61. Mean values for FSRQs and BL Lacs are -2.72 and -2.21.

\subsection{2-10 keV X-ray luminosity}
To do the fundamental analysis of Blazars, the X-ray luminosity in 2-10 keV is required. In this paper, the X-ray luminosity in 2-10 keV was calculated by the X-ray flux density from  NASA/IPAC Extragalactic Database (NED). In order to minimize errors caused by different observation devices, all data of X-ray flux density we choose are come from the same X-ray observation satellite as much as possible. Most of Blazars we collected can be observed by Swift with reliable data. Only a small number of samples did not have Swift observations, for these sources we collected the data of XMM-Netwon. Most data of flux density in 2-10 keV can be obtained from the NED. But there are still some data that do not have the direct observational data in the 2-10 keV. Because of the limited number of X-ray flux densities with direct observational data in the 2-10 keV range, we only collected another 0.3-10 keV flux density directly observed by Swift, and used the power-law model to estimate the 2-10 keV flux density to expand the sample and reduce the error to a certain extent. The X-ray flux densities in 0.3-10 keV of Blazars from NED can be converted into 2-10 keV by the power-law model with photon index $\tau$=1.87 for FSRQs \citep{Zhang2023b}. The photon index we chose is the average photon index of Blazars which is similar to \cite{Sahakyan2021}. Due to the small differences between two energy bands of X-rays, the uncertainty generated by the spectral index has little impact on the results.

\begin{eqnarray}
\nonumber
F_{\rm 2-10keV}=F_{\rm 0.3-10keV}\frac{\int_{\rm 2keV}^{\rm 10keV}\nu^{1-\tau}d\nu}{\int_{\rm 0.3keV}^{\rm 10keV}\nu^{1-\tau}d\nu}\\
=F_{\rm 0.3-10keV}\times0.5157,
\label{Eq3}
\end{eqnarray}
where $F_{\rm 2-10keV}$ and $F_{\rm 0.3-10keV}$ respectively are the X-ray flux density in 2-10 keV and 0.3-10 keV. $\nu$ is the energy dissipation of FSRQs. The 2-10keV X-ray luminosity distributions of the various classes are shown in part f of Figure \ref{fig:1}. The 2-10keV X-ray luminosity distributions for all blazars are $10^{42.5}$ $\sim$ $10^{47.5}$ erg $\rm{s^{-1}}$ and mean value is $10^{45.58\pm0.35}$ erg $\rm{s^{-1}}$. Mean values for FSRQs and BL Lacs are $10^{45.80\pm0.35}$ erg $\rm{s^{-1}}$ and $10^{44.85\pm0.38}$ erg $\rm{s^{-1}}$.

\subsection{Black hole spin energy}
The black hole spin energy was also considered as a variable for fundamental analysis.  However, due to the limitations of observation and estimation methods, it is difficult to measure the spin of large black holes directly by conventional calculation methods. We measured the $M_{\rm spin}$ of BH with the method of \cite{Daly2022}. The accuracy of this estimation method has been confirmed by previous work \citep{Daly2019} which can consistent with previous works such as the X-ray reflection method \citep{Miller2009,Reynolds2014}. To obtain the key data (Spin function F) of this method, the broad-line luminosity and the beam power of sample in this paper are required.

\subsubsection{\rm{Bolometric luminosity}}
In this paper, the bolometric luminosity $L_{\rm bol}$ of our sample were collected from the previous work by cross matching. $L_ {\rm BLR} $ which were collected from the previous work \citep{Xiong2014} can be converted into $L_{\rm bol}$  ($L_ {\rm bol} $= 10 $L_ {\rm BLR} $) with an average uncertainty of 0.3dex \citep{Calderone2013}. According to \cite{Celotti1997}, $L_{\rm BLR} $ usually were obtained by scaling several strong emission lines to the quasar template spectrum of \cite{Francis1991}, with $L_{\rm y\alpha}$ as a reference. The luminosity of blazars emission lines in the Sloan Digital Sky Survey (SDSS) DR7 quasar sample \citep{ Xiong2014} can be converted to the total luminosity of the broad lines \citep{Sbarrato2012, Celotti1997}. A proportions is set to estimate $L_ {\rm BLR} $ by these method that the $L_{\rm y\alpha}$ flux contribution is set to 100 and  the relative weights of H$\alpha$, H$\beta$, $\rm Mg_{II}$ and $\rm C_{IV}$ are 77, 22, 34 and 63, respectively. If there are more than one line was present, take the average of $L_ {\rm BLR} $ estimated from each line. The uncertainty of ${\rm log}L_{\rm {bol}}$ we took is 0.32 which is consistent with \cite{Calderone2013}. The $L_{\rm bol}$  distributions of the various classes are shown in part g of Figure \ref{fig:1}. The $L_{\rm bol}$  distributions for all blazars are $10^{42.5}$ $\sim$ $10^{49.5}$ erg $\rm{s^{-1}}$ and mean value is $10^{45.79\pm0.32}$ erg $\rm{s^{-1}}$. Mean values for FSRQs and BL Lacs are $10^{46.07\pm0.32}$ erg $\rm{s^{-1}}$ and $10^{44.85\pm0.32}$ erg $\rm{s^{-1}}$.

\subsubsection{\rm{Beam power}}
We used the theoretical model described by \cite{Willott1999} to calculate the beam power of our sample that replies on the relationships between beam power and radio luminosity at 151 MHz. For our sample, most of the 151 MHz radio luminosity directly comes from the Very Long Baseline Interferometry (VLBI) by cross-matching on NED. Only 6 FSRQ sources’ 151 MHz flux density has no direct observational data. Since the samples are all FSRQ, we collected the radio flux density at 140-160MHz for this sample and set the spectral index $\alpha=1$ to measure the 151 MHz radio flux density. Because 151 MHz radio flux density are taken from extended radio flux density of samples. The correction factor can be negligible because the frequencies (140-160 MHz) are approximate.
\begin{eqnarray}	
\nonumber
L_{\rm j} \approx 1.7 \times 10^{45}f^{3/2}{(\frac{L_{151}}{10^{44}{\rm erg~s}^{-1}})}^{6/7}{\rm erg~s}^{-1},\\
\label{Eq4}
\end{eqnarray}
where $L_{151}$ is the radio luminosity at 151 MHz in units of ${\rm erg~s}^{-1}$ and $L_{\rm j}$ is the beam power. \cite{Willott1999} contended that the normalization is highly uncertain, prompting the introduction of the factor f to accommodate these uncertainties. The value range of $f$ is 1$\leq f \leq$ 20 \citep{Cao2003,Godfrey2013,Wu2008,Fan2018, Fan2019}. According to analysis result of \cite{Daly2018} the uncertainty of beam power could be effected by the range of f. To reduce the uncertainty of our result as much as possible, we used the the lowest limit $f = 1$ to measured the beam power just like \cite{Cao2003}. The $L_{151}$ distributions of the various classes are shown in part h of Figure \ref{fig:1}. The $L_{151}$ distributions for all blazars are $10^{39.5}$ $\sim$ $10^{45}$ erg $\rm{s^{-1}}$ and mean value is $10^{42.85\pm0.03}$ erg $\rm{s^{-1}}$. Mean values for FSRQs and BL Lacs are $10^{43.17\pm0.04}$ erg $\rm{s^{-1}}$ and $10^{41.78\pm0.03}$ erg $\rm{s^{-1}}$.

\subsubsection{\rm{Spin function F}}
The spin function can be measured  by the following formula \citep{Daly2022,Zhang2023a} with beam power, bolometric luminosity and Eddington luminosity we got.
\begin{eqnarray}
\nonumber
\frac{f(j)}{f_{\rm max}}=(\frac{L_j}{g_jL_{\rm Edd}})(\frac{L_{\rm bol}}{g_{\rm bol}L_{\rm Edd}})^{-A},\\
\label{Eq5}
\end{eqnarray}
where j is BH spin, $L_{\rm bol}$ is the bolometric luminosity\citep{Netzer1990}, $L_{\rm Edd}$ is the Eddington luminosity which can be measured by BH mass, $g_i$ = 0.1 and $g_{bol}$ = 1 are directly obtained from \cite{Daly2019}, and A is a parameter defined by \citet{Daly2018}, it can be measured by the coefficient of fundamental plane of BH ($A=a/(a+b)$), where $a$ and $b$ are the value of parameters from \cite{Merloni2003}. In this paper, we set A as 0.43 \citep{Daly2018}. Based on the relationship among the black hole spin function F, the total mass M of BH (also known as $M_{\rm dyn}$) and the spin-mass energy ($M_{\rm spin}$) \citep{Blandford1990, Thorne1986, Rees1984, Misner1973, Zhang2023a}, we estimate $M_{\rm spin}$ by the black hole spin function $\rm{F^2}$.
M$\equiv$$M_{\rm dyn}$=$M_{\rm irr}$+$E_{\rm spin}$$c^{-2}$=$M_{\rm irr}$+$M_{\rm spin}$. \cite{Daly2022} showed that
\begin{eqnarray}
\nonumber
\frac{M_{\rm irr}}{M_{\rm dyn}}=(F^2+1)^{-1/2},\\
\frac{M_{\rm spin}}{M_{\rm dyn}}=1-\frac{M_{\rm irr}}{M_{\rm dyn}}=[1-(F^2+1)^{-1/2}],
\label{Eq6}
\end{eqnarray}
where F is the spin function, $F^2\equiv$f(j)/$f_{\rm max}$ \citep{Daly2022}, $M_{\rm dyn}$ is the dynamic black hole mass and $M_{\rm irr}$ is the irreducible black hole mass.

\begin{table*}[htbp]
\centering
\captionsetup{labelsep=none}
\caption{The fundamental plane of Blazars}
\label{table:1}
\begin{tabular}{cccccccc}
\toprule
    ${\rm [1]}$ & ${\rm [2]}$ & ${\rm [3]}$ & ${\rm [4]}$ & ${\rm [5]}$ & ${\rm [6]}$ & ${\rm [7]}$  & ${\rm [8]}$\\
    ${\rm name}$ & type  & z & ${\rm log}M_{\rm dyn}$ & ${\rm log}L_{\rm X}$ & ${\rm log}L_{\rm R}$ &  ${\rm log}f_{\rm b}$  &  ${\rm log R}$\\
    \midrule
    0106+013/4C+01.02	&	FSRQ	&	2.099	&	8.830 	$\pm$	0.300 	&	46.460 	$\pm$	0.415 	&	45.234 	$\pm$	0.003 	&	-3.208 	&	3.991 	\\
    0110+495	&	FSRQ	&	0.389	&	8.340 	$\pm$	0.300 	&	44.567 	$\pm$	0.370 	&	42.966 	$\pm$	0.102 	&	-2.221 	&	3.536 	\\
    0133+476/OC 457	&	FSRQ	&	0.859	&	8.515 	$\pm$	0.300 	&	45.623 	$\pm$	0.414 	&	44.462 	$\pm$	0.017 	&	-2.618 	&	3.752 	\\
    0135-247/PKS0135-247	&	FSRQ	&	0.835	&	9.120 	$\pm$	0.300 	&	45.695 	$\pm$	0.404 	&	44.131 	$\pm$	0.022 	&	-2.623 	&	4.200 	\\
    0234+285/4C +28.07	&	FSRQ	&	1.213	&	9.220 	$\pm$	0.300 	&	46.064 	$\pm$	0.409 	&	44.642 	$\pm$	0.007 	&	-2.516 	&	3.502 	\\
    0235+164/PKS 0235+164	&	Bllac	&	0.940	&	9.073 	$\pm$	0.300 	&	45.882 	$\pm$	0.426 	&	44.005 	$\pm$	0.034 	&	-2.467 	&	2.966 	\\
    0336-019/PKS 0336-01	&	FSRQ	&	0.852	&	8.935 	$\pm$	0.300 	&	45.503 	$\pm$	0.413 	&	44.434 	$\pm$	0.005 	&	-3.032 	&	3.584 	\\
    0403-132/PKS 0403-13	&	FSRQ	&	0.571	&	9.075 	$\pm$	0.300 	&	45.605 	$\pm$	0.320 	&	44.108 	$\pm$	0.016 	&	-2.703 	&	3.616 	\\
    0420-014/PKS 0420-01	&	FSRQ	&	0.916	&	8.813 	$\pm$	0.300 	&	45.827 	$\pm$	0.434 	&	44.663 	$\pm$	0.003 	&	-2.400 	&	3.196 	\\
    0458-020/PKS 0458-020	&	FSRQ	&	2.291	&	8.965 	$\pm$	0.300 	&	46.593 	$\pm$	0.408 	&	45.284 	$\pm$	0.011 	&	-2.741 	&	3.996 	\\
    0528+134	&	FSRQ	&	2.060	&	9.800 	$\pm$	0.300 	&	46.604 	$\pm$	0.423 	&	45.031 	$\pm$	0.027 	&	-2.962 	&	4.172 	\\
    OG 050	&	Bllac	&	1.254	&	8.430 	$\pm$	0.350 	&	45.989 	$\pm$	0.408 	&	44.587 	$\pm$	0.026 	&	-2.886 	&	3.739 	\\
    0537-441/PKS 0537-441	&	Bllac	&	0.892	&	8.565 	$\pm$	0.300 	&	46.173 	$\pm$	0.032 	&	44.576 	$\pm$	0.022 	&	-2.829 	&	3.086 	\\
    0605-085/PKS 0605-085	&	FSRQ	&	0.872	&	8.628 	$\pm$	0.300 	&	45.648 	$\pm$	0.406 	&	44.357 	$\pm$	0.022 	&	-3.009 	&	3.087 	\\
    0637-752/PKS 0637-75	&	FSRQ	&	0.653	&	9.110 	$\pm$	0.300 	&	45.782 	$\pm$	0.421 	&	44.404 	$\pm$	0.022 	&	-2.782 	&	3.482 	\\
    B3 0650+453	&	FSRQ	&	0.928	&	8.170 	$\pm$	0.340 	&	45.248 	$\pm$	0.377 	&	43.369 	$\pm$	0.239 	&	-2.475 	&	3.340 	\\
    0716+714	&	Bllac	&	0.300	&	8.100 	$\pm$	0.300 	&	45.269 	$\pm$	0.431 	&	43.151 	$\pm$	0.087 	&	-2.319 	&	2.316 	\\
    0735+178	&	Bllac	&	0.424	&	8.300 	$\pm$	0.300 	&	44.499 	$\pm$	0.399 	&	43.145 	$\pm$	0.061 	&	-2.504 	&	3.058 	\\
    0736+017/PKS 0736+01	&	FSRQ	&	0.189	&	8.110 	$\pm$	0.300 	&	44.468 	$\pm$	0.424 	&	42.864 	$\pm$	0.052 	&	-2.736 	&	3.111 	\\
    0748+126/PKS 0748+126	&	FSRQ	&	0.889	&	8.150 	$\pm$	0.300 	&	45.837 	$\pm$	0.409 	&	44.524 	$\pm$	0.015 	&	-2.675 	&	3.541 	\\
    0754+100	&	Bllac	&	0.266	&	8.200 	$\pm$	0.300 	&	44.731 	$\pm$	0.418 	&	42.865 	$\pm$	0.042 	&	-2.974 	&	2.514 	\\
    PKS0808+019	&	Bllac	&	1.148	&	8.500 	$\pm$	0.300 	&	45.265 	$\pm$	0.366 	&	44.388 	$\pm$	0.035 	&	-2.615 	&	3.583 	\\
    4C +39.23	&	FSRQ	&	1.216	&	8.550 	$\pm$	0.300 	&	45.544 	$\pm$	0.326 	&	44.245 	$\pm$	0.068 	&	-2.779 	&	3.046 	\\
    OJ 535	&	FSRQ	&	1.418	&	9.260 	$\pm$	0.300 	&	45.902 	$\pm$	0.378 	&	44.375 	$\pm$	0.054 	&	-2.822 	&	3.187 	\\
    0823+033/PKS 0823+033	&	Bllac	&	0.506	&	8.830 	$\pm$	0.300 	&	44.967 	$\pm$	0.407 	&	43.699 	$\pm$	0.058 	&	-2.486 	&	2.641 	\\
    B2 0827+24	&	FSRQ	&	0.940	&	8.953 	$\pm$	0.300 	&	46.004 	$\pm$	0.424 	&	44.087 	$\pm$	0.039 	&	-3.104 	&	3.024 	\\
    PKS 0829+046	&	Bllac	&	0.174	&	8.630 	$\pm$	0.300 	&	43.919 	$\pm$	0.415 	&	42.403 	$\pm$	0.073 	&	-2.111 	&	2.126 	\\
    OJ 451	&	FSRQ	&	0.249	&	9.680 	$\pm$	0.090 	&	43.805 	$\pm$	0.315 	&	42.252 	$\pm$	0.087 	&	-1.972 	&	2.367 	\\
    0836+710/4C+71.07	&	FSRQ	&	2.172	&	9.360 	$\pm$	0.300 	&	47.483 	$\pm$	0.012 	&	44.950 	$\pm$	0.031 	&	-3.198 	&	2.859 	\\
    0851+202/OJ 287	&	Bllac	&	0.306	&	8.563 	$\pm$	0.300 	&	45.009 	$\pm$	0.431 	&	43.418 	$\pm$	0.022 	&	-2.229 	&	2.792 	\\
    S4 0859+470	&	FSRQ	&	1.466	&	9.250 	$\pm$	0.030 	&	45.632 	$\pm$	0.320 	&	44.529 	$\pm$	0.011 	&	-2.863 	&	3.645 	\\
    0906+015/PKS 0906+01	&	FSRQ	&	1.024	&	9.003 	$\pm$	0.300 	&	45.871 	$\pm$	0.415 	&	44.115 	$\pm$	0.056 	&	-2.617 	&	2.889 	\\
    S4 0913+39	&	FSRQ	&	1.267	&	8.620 	$\pm$	0.150 	&	45.193 	$\pm$	0.320 	&	43.968 	$\pm$	0.074 	&	-2.727 	&	3.237 	\\
    B3 0917+449	&	FSRQ	&	2.190	&	9.283 	$\pm$	0.300 	&	46.758 	$\pm$	0.406 	&	44.740 	$\pm$	0.051 	&	-2.893 	&	2.792 	\\
    OK 630	&	FSRQ	&	1.453	&	8.930 	$\pm$	0.160 	&	45.819 	$\pm$	0.375 	&	44.557 	$\pm$	0.037 	&	-2.760 	&	3.678 	\\
    4C +40.24	&	FSRQ	&	1.249	&	8.950 	$\pm$	0.060 	&	45.717 	$\pm$	0.375 	&	44.487 	$\pm$	0.074 	&	-3.275 	&	3.471 	\\
    0953+254/B2 0954+25A	&	FSRQ	&	0.707	&	8.876 	$\pm$	0.300 	&	45.256 	$\pm$	0.405 	&	43.602 	$\pm$	0.099 	&	-2.555 	&	3.252 	\\
    4C+55.17	&	FSRQ	&	0.896	&	8.338 	$\pm$	0.300 	&	45.159 	$\pm$	0.023 	&	43.764 	$\pm$	0.020 	&	-2.800 	&	2.848 	\\
    4C+23.24	&	FSRQ	&	0.566	&	8.510 	$\pm$	0.098 	&	45.098 	$\pm$	0.362 	&	43.636 	$\pm$	0.079 	&	-2.483 	&	3.117 	\\
    1ES 1011+496	&	Bllac	&	0.212	&	8.300 	$\pm$	0.300 	&	45.398 	$\pm$	0.429 	&	42.015 	$\pm$	0.259 	&	-1.987 	&	1.627 	\\
    1055+018/PKS 1055+01	&	Bllac	&	0.888	&	8.410 	$\pm$	0.300 	&	45.881 	$\pm$	0.422 	&	44.605 	$\pm$	0.015 	&	-2.190 	&	3.660 	\\
    Mkn 421	&	Bllac	&	0.03	&	8.153 	$\pm$	0.300 	&	44.990 	$\pm$	0.320 	&	40.399 	$\pm$	0.066 	&	-1.459 	&	0.060 	\\
    1127-145	&	FSRQ	&	1.184	&	9.180 	$\pm$	0.300 	&	46.657 	$\pm$	0.017 	&	44.975 	$\pm$	0.039 	&	-2.984 	&	3.365 	\\
    4C+29.45	&	FSRQ	&	0.724	&	8.499 	$\pm$	0.300 	&	45.458 	$\pm$	0.419 	&	43.911 	$\pm$	0.045 	&	-3.100 	&	3.323 	\\
    1202-262/PKS 1203-26	&	FSRQ	&	0.789	&	8.795 	$\pm$	0.300 	&	45.602 	$\pm$	0.320 	&	43.937 	$\pm$	0.015 	&	-2.660 	&	3.982 	\\
    1219+285	&	Bllac	&	0.102	&	7.700 	$\pm$	0.300 	&	43.606 	$\pm$	0.425 	&	41.563 	$\pm$	0.163 	&	-1.850 	&	1.514 	\\
    4C+04.42	&	FSRQ	&	0.965	&	8.305 	$\pm$	0.300 	&	45.956 	$\pm$	0.409 	&	43.754 	$\pm$	0.078 	&	-2.607 	&	1.702 	\\
    4C+21.35	&	FSRQ	&	0.432	&	8.737 	$\pm$	0.300 	&	45.263 	$\pm$	0.320 	&	43.225 	$\pm$	0.116 	&	-3.609 	&	2.580 	\\
    1226+023/3C 273	&	FSRQ	&	0.158	&	8.510 	$\pm$	0.300 	&	45.723 	$\pm$	0.430 	&	44.087 	$\pm$	0.005 	&	-2.581 	&	2.580 	\\
    1253-055/3C 279	&	FSRQ	&	0.536	&	8.503 	$\pm$	0.300 	&	45.922 	$\pm$	0.431 	&	44.660 	$\pm$	0.005 	&	-2.937 	&	3.451 	\\
    1308+326/B2 1308+32	&	FSRQ	&	0.996	&	8.478 	$\pm$	0.300 	&	45.823 	$\pm$	0.414 	&	44.185 	$\pm$	0.018 	&	-2.986 	&	3.982 	\\
    B2 1315+34A	&	FSRQ	&	1.056	&	9.215 	$\pm$	0.300 	&	45.208 	$\pm$	0.308 	&	43.556 	$\pm$	0.200 	&	-2.568 	&	3.464 	\\
    B2 1324+22	&	FSRQ	&	1.4	&	9.245 	$\pm$	0.300 	&	46.011 	$\pm$	0.150 	&	44.426 	$\pm$	0.047 	&	-2.376 	&	3.033 	\\
    1334-127/PKS 1335-127	&	FSRQ	&	0.539	&	7.980 	$\pm$	0.300 	&	45.580 	$\pm$	0.416 	&	44.195 	$\pm$	0.018 	&	-2.611 	&	3.516 	\\
    B3 1343+451	&	FSRQ	&	2.534	&	8.980 	$\pm$	0.990 	&	46.398 	$\pm$	0.043 	&	44.363 	$\pm$	0.157 	&	-2.733 	&	3.543 	\\
    S5 1357+76	&	FSRQ	&	1.585	&	8.255 	$\pm$	0.210 	&	45.409 	$\pm$	0.339 	&	44.106 	$\pm$	0.123 	&	-2.726 	&	3.501 	\\
    B3 1417+385	&	FSRQ	&	1.831	&	8.620 	$\pm$	0.250 	&	45.799 	$\pm$	0.387 	&	44.713 	$\pm$	0.039 	&	-2.759 	&	3.639 	\\
    PKS 1434+235	&	FSRQ	&	1.548	&	8.375 	$\pm$	0.300 	&	45.776 	$\pm$	0.324 	&	44.237 	$\pm$	0.087 	&	-2.756 	&	3.739 	\\
    PKS 1502+106	&	FSRQ	&	1.839	&	9.107 	$\pm$	0.300 	&	46.221 	$\pm$	0.410 	&	44.609 	$\pm$	0.051 	&	-2.665 	&	3.355 	\\

\bottomrule
\end{tabular}
\begin{flushleft}
Note: The first column is name; the second column is type of blazars; the third column is redshift; the fourth column is the Logarithm of Black hole dynamical mass; the fifth column is Logarithm of x-ray luminosity in 2-10 keV (in units of ${\rm erg~s^{-1}}$); the sixth column is Logarithm of 5 GHz  radio luminosity (in units of ${\rm erg~s^{-1}}$); the seventh column is Logarithm of the beaming factor; the eighth column is Logarithm of the radio loudness.
\end{flushleft}
\end{table*}

\begin{table*}
\centering
\ContinuedFloat 
\captionsetup{labelsep=none} 
\caption{continued}
\begin{tabular}{cccccccc}
\toprule
    ${\rm [1]}$ & ${\rm [2]}$ & ${\rm [3]}$ & ${\rm [4]}$ & ${\rm [5]}$ & ${\rm [6]}$ & ${\rm [7]}$  & ${\rm [8]}$\\
    ${\rm name}$ & type  & z & ${\rm log}M_{\rm dyn}$ & ${\rm log}L_{\rm X}$ & ${\rm log}L_{\rm R}$ &  ${\rm log}f_{\rm b}$  &  ${\rm log R}$\\
    \midrule
    1508-055	&	FSRQ	&	1.185	&	8.970 	$\pm$	0.300 	&	45.578 	$\pm$	0.035 	&	44.606 	$\pm$	0.022 	&	-2.912 	&	3.335 	\\
    1510-089/PKS 1510-08	&	FSRQ	&	0.361	&	8.363 	$\pm$	0.300 	&	45.426 	$\pm$	0.028 	&	43.709 	$\pm$	0.031 	&	-2.929 	&	3.259 	\\
    1514-241	&	Bllac	&	0.049	&	7.650 	$\pm$	0.300 	&	43.224 	$\pm$	0.419 	&	41.912 	$\pm$	0.022 	&	-1.776 	&	3.336 	\\
    B2 1520+31	&	FSRQ	&	1.484	&	8.920 	$\pm$	0.310 	&	45.506 	$\pm$	0.342 	&	44.089 	$\pm$	0.113 	&	-2.606 	&	1.344 	\\
    1546+027/PKS 1546+027	&	FSRQ	&	0.414	&	8.618 	$\pm$	0.300 	&	44.918 	$\pm$	0.422 	&	43.490 	$\pm$	0.035 	&	-2.336 	&	3.628 	\\
    4C+05.64	&	FSRQ	&	1.422	&	9.180 	$\pm$	0.300 	&	45.972 	$\pm$	0.386 	&	44.908 	$\pm$	0.090 	&	-2.897 	&	3.081 	\\
    4C+10.45	&	FSRQ	&	1.226	&	8.970 	$\pm$	0.300 	&	45.900 	$\pm$	0.434 	&	44.316 	$\pm$	0.039 	&	-2.854 	&	4.001 	\\
    1611+343/B2 1611+34	&	FSRQ	&	1.397	&	9.343 	$\pm$	0.300 	&	45.899 	$\pm$	0.405 	&	44.847 	$\pm$	0.018 	&	-2.108 	&	3.421 	\\
    1622-297	&	FSRQ	&	0.815	&	9.100 	$\pm$	0.300 	&	45.408 	$\pm$	0.320 	&	44.228 	$\pm$	0.026 	&	-2.721 	&	3.371 	\\
    1633+382/4C+38.41	&	FSRQ	&	1.814	&	9.369 	$\pm$	0.300 	&	46.718 	$\pm$	0.427 	&	45.090 	$\pm$	0.024 	&	-3.287 	&	2.910 	\\
    4C +47.44	&	FSRQ	&	0.735	&	8.565 	$\pm$	0.475 	&	45.330 	$\pm$	0.371 	&	43.655 	$\pm$	0.077 	&	-2.564 	&	3.455 	\\
    Mkn 501	&	Bllac	&	0.034	&	8.837 	$\pm$	0.300 	&	44.210 	$\pm$	0.431 	&	41.236 	$\pm$	0.053 	&	-1.613 	&	3.596 	\\
    B3 1708+433	&	FSRQ	&	1.027	&	7.920 	$\pm$	0.350 	&	45.110 	$\pm$	0.314 	&	43.264 	$\pm$	0.372 	&	-2.350 	&	0.819 	\\
    1726+455	&	FSRQ	&	0.717	&	8.220 	$\pm$	0.300 	&	45.054 	$\pm$	0.395 	&	43.826 	$\pm$	0.050 	&	-2.532 	&	2.870 	\\
    1730-130	&	FSRQ	&	0.902	&	9.300 	$\pm$	0.300 	&	45.615 	$\pm$	0.418 	&	44.851 	$\pm$	0.007 	&	-3.934 	&	3.013 	\\
    1739+522	&	FSRQ	&	1.375	&	9.320 	$\pm$	0.300 	&	46.028 	$\pm$	0.402 	&	44.404 	$\pm$	0.086 	&	-2.718 	&	4.961 	\\
    4C+09.57/PKS1749+096	&	Bllac	&	0.322	&	8.413 	$\pm$	0.300 	&	45.019 	$\pm$	0.422 	&	43.449 	$\pm$	0.026 	&	-2.097 	&	2.845 	\\
    CGRaBS J1800+7828/S5 1803+78	&	Bllac	&	0.68	&	8.260 	$\pm$	0.300 	&	45.498 	$\pm$	0.419 	&	44.133 	$\pm$	0.022 	&	-2.248 	&	3.179 	\\
    3C 371	&	Bllac	&	0.051	&	8.605 	$\pm$	0.300 	&	42.956 	$\pm$	0.423 	&	41.524 	$\pm$	0.033 	&	-1.775 	&	3.092 	\\
    1823+568/4C+56.27	&	Bllac	&	0.664	&	9.260 	$\pm$	0.300 	&	45.458 	$\pm$	0.407 	&	43.763 	$\pm$	0.016 	&	-2.319 	&	1.144 	\\
    1849+670/CGRaBS J1849+6705	&	FSRQ	&	0.657	&	9.140 	$\pm$	0.300 	&	45.270 	$\pm$	0.411 	&	43.845 	$\pm$	0.040 	&	-2.412 	&	3.047 	\\
    1921-293/PKS B1921-293	&	FSRQ	&	0.353	&	8.695 	$\pm$	0.300 	&	45.320 	$\pm$	0.414 	&	44.341 	$\pm$	0.005 	&	-2.729 	&	3.085 	\\
    2145+067/4C+06.69	&	FSRQ	&	0.99	&	8.870 	$\pm$	0.300 	&	46.225 	$\pm$	0.422 	&	44.892 	$\pm$	0.008 	&	-2.097 	&	4.042 	\\
    2155-152/PKS 2155-152	&	FSRQ	&	0.672	&	7.590 	$\pm$	0.300 	&	45.307 	$\pm$	0.414 	&	44.237 	$\pm$	0.022 	&	-2.703 	&	3.250 	\\
    BL Lac	&	Bllac	&	0.069	&	8.210 	$\pm$	0.300 	&	43.828 	$\pm$	0.320 	&	42.048 	$\pm$	0.030 	&	-1.767 	&	3.861 	\\
    PKS 2209+236	&	FSRQ	&	1.125	&	8.460 	$\pm$	0.450 	&	45.217 	$\pm$	0.398 	&	44.150 	$\pm$	0.058 	&	-2.597 	&	2.265 	\\
    2223-052/3C 446	&	FSRQ	&	1.404	&	8.417 	$\pm$	0.300 	&	43.176 	$\pm$	0.320 	&	45.186 	$\pm$	0.011 	&	-2.636 	&	4.115 	\\
    PKS 2227-08	&	FSRQ	&	1.56	&	8.785 	$\pm$	0.300 	&	46.420 	$\pm$	0.402 	&	44.845 	$\pm$	0.014 	&	-2.301 	&	3.712 	\\
    2230+114/CTA 102	&	FSRQ	&	1.037	&	8.780 	$\pm$	0.300 	&	46.268 	$\pm$	0.320 	&	44.937 	$\pm$	0.036 	&	-2.676 	&	3.331 	\\
    2251+158/3C 454.3	&	FSRQ	&	0.859	&	8.825 	$\pm$	0.300 	&	46.884 	$\pm$	0.320 	&	45.013 	$\pm$	0.013 	&	-2.881 	&	3.677 	\\
    TXS 2331+073	&	FSRQ	&	0.401	&	8.370 	$\pm$	0.500 	&	44.780 	$\pm$	0.408 	&	43.255 	$\pm$	0.023 	&	-2.281 	&	3.495 	\\
    2345-16/PKS 2345-16	&	FSRQ	&	0.576	&	8.595 	$\pm$	0.300 	&	45.369 	$\pm$	0.413 	&	44.007 	$\pm$	0.022 	&	-2.631 	&	2.974 	\\
\bottomrule
\end{tabular}
\begin{flushleft}
Note: The first column is name; the second column is type of blazars; the third column is redshift; the fourth column is the Logarithm of Black hole dynamical mass; the fifth column is Logarithm of x-ray luminosity in 2-10 keV (in units of ${\rm erg~s^{-1}}$); the sixth column is Logarithm of 5 GHz  radio luminosity (in units of ${\rm erg~s^{-1}}$); the seventh column is Logarithm of the beaming factor; the eighth column is Logarithm of the radio loudness.
\end{flushleft}
\end{table*}

\begin{table*}
\centering
\captionsetup{labelsep=none}
\caption{The BH spin energy of Blazars}
\label{table:2}
\begin{tabular}{cccccccc}
\toprule
${\rm [1]}$ & ${\rm [2]}$ & ${\rm [3]}$ & ${\rm [4]}$ & ${\rm [5]}$ & ${\rm [6]}$ & ${\rm [7]}$ & ${\rm [8]}$ \\
 ${\rm name}$ & ${\rm log}L_{\rm {151}}$ & ${\rm log}L_{\rm j}$ & ${\rm log}L_{\rm {Edd}}$ & ${\rm log}L_{\rm {bol}}$ & ${\rm log}{\rm F}$ & ${\rm log}M_{\rm spin}$ & ${\rm log}M_{\rm irr}$ \\
\midrule
0106+013/4C+01.02	&	44.378 	$\pm$	0.036 	&	45.171 	$\pm$	0.031 	&	46.944 	$\pm$	0.300 	&	47.135 	&	-0.428 	$\pm$	0.129 	&	8.767 	$\pm$	0.243 	&	7.962 	$\pm$	-0.613 	\\
0110+495	&	42.027 	$\pm$	0.036 	&	43.540 	$\pm$	0.031 	&	46.454 	$\pm$	0.300 	&	45.780 	&	-0.812 	$\pm$	0.129 	&	8.335 	$\pm$	0.285 	&	6.407 	$\pm$	-1.155 	\\
0133+476/OC 457	&	42.877 	$\pm$	0.022 	&	44.268 	$\pm$	0.019 	&	46.629 	$\pm$	0.300 	&	45.444 	&	-0.426 	$\pm$	0.141 	&	8.486 	$\pm$	0.263 	&	7.320 	$\pm$	-0.790 	\\
0135-247/PKS0135-247	&	43.415 	$\pm$	0.036 	&	44.729 	$\pm$	0.031 	&	47.234 	$\pm$	0.300 	&	46.343 	&	-0.561 	$\pm$	0.129 	&	9.104 	$\pm$	0.273 	&	7.673 	$\pm$	-0.912 	\\
0234+285/4C +28.07	&	43.451 	$\pm$	0.036 	&	44.760 	$\pm$	0.031 	&	47.334 	$\pm$	0.300 	&	46.244 	&	-0.553 	$\pm$	0.129 	&	9.204 	$\pm$	0.272 	&	7.789 	$\pm$	-0.905 	\\
0235+164/PKS 0235+164	&	42.839 	$\pm$	0.036 	&	44.235 	$\pm$	0.031 	&	47.187 	$\pm$	0.300 	&	44.920 	&	-0.489 	$\pm$	0.129 	&	9.052 	$\pm$	0.268 	&	7.762 	$\pm$	-0.844 	\\
0336-019/PKS 0336-01	&	42.903 	$\pm$	0.036 	&	44.290 	$\pm$	0.031 	&	47.049 	$\pm$	0.300 	&	46.000 	&	-0.654 	$\pm$	0.129 	&	8.925 	$\pm$	0.278 	&	7.310 	$\pm$	-1.001 	\\
0403-132/PKS 0403-13	&	43.272 	$\pm$	0.036 	&	44.607 	$\pm$	0.031 	&	47.189 	$\pm$	0.300 	&	46.250 	&	-0.589 	$\pm$	0.129 	&	9.061 	$\pm$	0.274 	&	7.575 	$\pm$	-0.939 	\\
0420-014/PKS 0420-01	&	43.049 	$\pm$	0.036 	&	44.415 	$\pm$	0.031 	&	46.927 	$\pm$	0.300 	&	45.901 	&	-0.535 	$\pm$	0.129 	&	8.796 	$\pm$	0.271 	&	7.415 	$\pm$	-0.888 	\\
0458-020/PKS 0458-020	&	44.385 	$\pm$	0.036 	&	45.560 	$\pm$	0.031 	&	47.079 	$\pm$	0.300 	&	46.304 	&	-0.093 	$\pm$	0.129 	&	8.856 	$\pm$	0.224 	&	8.312 	$\pm$	-0.492 	\\
0528+134	&	43.443 	$\pm$	0.036 	&	44.753 	$\pm$	0.031 	&	47.914 	$\pm$	0.300 	&	49.650 	&	-1.454 	$\pm$	0.129 	&	9.800 	$\pm$	0.296 	&	6.591 	$\pm$	-1.790 	\\
OG 050	&	43.407 	$\pm$	0.036 	&	44.722 	$\pm$	0.031 	&	46.544 	$\pm$	0.350 	&	45.864 	&	-0.265 	$\pm$	0.129 	&	8.374 	$\pm$	0.298 	&	7.515 	$\pm$	-0.385 	\\
0537-441/PKS 0537-441	&	43.297 	$\pm$	0.036 	&	44.628 	$\pm$	0.031 	&	46.679 	$\pm$	0.300 	&	46.020 	&	-0.384 	$\pm$	0.129 	&	8.531 	$\pm$	0.259 	&	7.445 	$\pm$	-0.746 	\\
0605-085/PKS 0605-085	&	43.257 	$\pm$	0.036 	&	44.593 	$\pm$	0.031 	&	46.741 	$\pm$	0.300 	&	45.600 	&	-0.329 	$\pm$	0.129 	&	8.584 	$\pm$	0.254 	&	7.604 	$\pm$	-0.695 	\\
0637-752/PKS 0637-75	&	43.737 	$\pm$	0.036 	&	45.005 	$\pm$	0.031 	&	47.224 	$\pm$	0.300 	&	46.228 	&	-0.395 	$\pm$	0.129 	&	9.077 	$\pm$	0.260 	&	7.969 	$\pm$	-0.757 	\\
B3 0650+453	&	42.552 	$\pm$	0.057 	&	43.989 	$\pm$	0.049 	&	46.284 	$\pm$	0.340 	&	45.260 	&	-0.427 	$\pm$	0.111 	&	8.142 	$\pm$	0.303 	&	6.972 	$\pm$	-0.655 	\\
0716+714	&	41.993 	$\pm$	0.007 	&	43.510 	$\pm$	0.006 	&	46.214 	$\pm$	0.300 	&	45.766 	&	-0.756 	$\pm$	0.154 	&	8.093 	$\pm$	0.283 	&	6.278 	$\pm$	-1.112 	\\
0735+178	&	42.386 	$\pm$	0.036 	&	43.847 	$\pm$	0.031 	&	46.414 	$\pm$	0.300 	&	47.600 	&	-1.038 	$\pm$	0.129 	&	8.298 	$\pm$	0.291 	&	5.920 	$\pm$	-1.378 	\\
0736+017/PKS 0736+01	&	41.635 	$\pm$	0.036 	&	43.203 	$\pm$	0.031 	&	46.224 	$\pm$	0.300 	&	45.195 	&	-0.789 	$\pm$	0.129 	&	8.104 	$\pm$	0.284 	&	6.223 	$\pm$	-1.133 	\\
0748+126/PKS 0748+126	&	43.010 	$\pm$	0.036 	&	44.382 	$\pm$	0.031 	&	46.264 	$\pm$	0.300 	&	45.950 	&	-0.374 	$\pm$	0.129 	&	8.114 	$\pm$	0.258 	&	7.048 	$\pm$	-0.736 	\\
0754+100	&	41.417 	$\pm$	0.036 	&	43.017 	$\pm$	0.031 	&	46.314 	$\pm$	0.300 	&	47.500 	&	-1.404 	$\pm$	0.129 	&	8.200 	$\pm$	0.296 	&	5.091 	$\pm$	-1.740 	\\
PKS0808+019	&	42.817 	$\pm$	0.036 	&	44.216 	$\pm$	0.031 	&	46.614 	$\pm$	0.300 	&	44.623 	&	-0.271 	$\pm$	0.129 	&	8.445 	$\pm$	0.247 	&	7.574 	$\pm$	-0.644 	\\
4C +39.23	&	43.760 	$\pm$	0.016 	&	45.025 	$\pm$	0.013 	&	46.664 	$\pm$	0.300 	&	45.834 	&	-0.141 	$\pm$	0.147 	&	8.459 	$\pm$	0.232 	&	7.828 	$\pm$	-0.540 	\\
OJ 535	&	43.615 	$\pm$	0.017 	&	44.900 	$\pm$	0.015 	&	47.374 	$\pm$	0.300 	&	46.302 	&	-0.506 	$\pm$	0.146 	&	9.240 	$\pm$	0.269 	&	7.916 	$\pm$	-0.868 	\\
0823+033/PKS 0823+033	&	42.039 	$\pm$	0.036 	&	43.550 	$\pm$	0.031 	&	46.944 	$\pm$	0.300 	&	44.374 	&	-0.644 	$\pm$	0.129 	&	8.819 	$\pm$	0.277 	&	7.224 	$\pm$	-0.992 	\\
B2 0827+24	&	42.679 	$\pm$	0.036 	&	44.098 	$\pm$	0.031 	&	47.066 	$\pm$	0.300 	&	45.989 	&	-0.752 	$\pm$	0.129 	&	8.946 	$\pm$	0.282 	&	7.136 	$\pm$	-1.097 	\\
PKS 0829+046	&	41.317 	$\pm$	0.036 	&	42.931 	$\pm$	0.031 	&	46.744 	$\pm$	0.300 	&	43.568 	&	-0.724 	$\pm$	0.129 	&	8.622 	$\pm$	0.281 	&	6.870 	$\pm$	-1.069 	\\
OJ 451	&	41.105 	$\pm$	0.069 	&	42.749 	$\pm$	0.059 	&	47.794 	$\pm$	0.090 	&	44.070 	&	-1.222 	$\pm$	0.101 	&	9.679 	$\pm$	0.083 	&	6.934 	$\pm$	-2.677 	\\
0836+710/4C+71.07	&	44.378 	$\pm$	0.008 	&	45.555 	$\pm$	0.007 	&	47.474 	$\pm$	0.300 	&	47.430 	&	-0.450 	$\pm$	0.153 	&	9.334 	$\pm$	0.266 	&	8.120 	$\pm$	-0.819 	\\
0851+202/OJ 287	&	41.489 	$\pm$	0.036 	&	43.078 	$\pm$	0.031 	&	46.677 	$\pm$	0.300 	&	44.580 	&	-0.849 	$\pm$	0.129 	&	8.559 	$\pm$	0.286 	&	6.559 	$\pm$	-1.191 	\\
S4 0859+470	&	43.945 	$\pm$	0.010 	&	45.183 	$\pm$	0.009 	&	47.364 	$\pm$	0.030 	&	46.264 	&	-0.354 	$\pm$	0.151 	&	9.211 	$\pm$	-0.021 	&	8.183 	$\pm$	-0.389 	\\
0906+015/PKS 0906+01	&	42.962 	$\pm$	0.036 	&	44.341 	$\pm$	0.031 	&	47.117 	$\pm$	0.300 	&	46.100 	&	-0.669 	$\pm$	0.129 	&	8.994 	$\pm$	0.279 	&	7.349 	$\pm$	-1.017 	\\
S4 0913+39	&	43.498 	$\pm$	0.021 	&	44.800 	$\pm$	0.018 	&	46.734 	$\pm$	0.150 	&	45.804 	&	-0.267 	$\pm$	0.143 	&	8.564 	$\pm$	0.092 	&	7.701 	$\pm$	-0.339 	\\
B3 0917+449	&	43.867 	$\pm$	0.034 	&	45.117 	$\pm$	0.029 	&	47.397 	$\pm$	0.300 	&	46.850 	&	-0.523 	$\pm$	0.132 	&	9.265 	$\pm$	0.270 	&	7.909 	$\pm$	-0.877 	\\
OK 630	&	43.334 	$\pm$	0.015 	&	44.660 	$\pm$	0.013 	&	47.044 	$\pm$	0.160 	&	46.055 	&	-0.479 	$\pm$	0.147 	&	8.907 	$\pm$	0.125 	&	7.636 	$\pm$	-0.691 	\\
4C +40.24	&	43.615 	$\pm$	0.036 	&	44.900 	$\pm$	0.031 	&	47.064 	$\pm$	0.060 	&	46.502 	&	-0.461 	$\pm$	0.129 	&	8.925 	$\pm$	0.020 	&	7.690 	$\pm$	-0.459 	\\
0953+254/B2 0954+25A	&	42.311 	$\pm$	0.061 	&	43.783 	$\pm$	0.053 	&	46.990 	$\pm$	0.300 	&	45.950 	&	-0.880 	$\pm$	0.108 	&	8.873 	$\pm$	0.286 	&	6.810 	$\pm$	-1.211 	\\
4C+55.17	&	43.535 	$\pm$	0.007 	&	44.832 	$\pm$	0.006 	&	46.451 	$\pm$	0.300 	&	45.574 	&	-0.121 	$\pm$	0.155 	&	8.239 	$\pm$	0.230 	&	7.644 	$\pm$	-0.527 	\\
4C+23.24	&	42.771 	$\pm$	0.020 	&	44.177 	$\pm$	0.017 	&	46.623 	$\pm$	0.098 	&	45.893 	&	-0.566 	$\pm$	0.143 	&	8.494 	$\pm$	0.067 	&	7.053 	$\pm$	-0.744 	\\
1ES 1011+496	&	41.067 	$\pm$	0.028 	&	42.717 	$\pm$	0.024 	&	46.414 	$\pm$	0.300 	&	46.505 	&	-1.368 	$\pm$	0.136 	&	8.300 	$\pm$	0.296 	&	5.262 	$\pm$	-1.708 	\\
1055+018/PKS 1055+01	&	43.494 	$\pm$	0.036 	&	44.796 	$\pm$	0.031 	&	46.524 	$\pm$	0.300 	&	45.509 	&	-0.146 	$\pm$	0.129 	&	8.320 	$\pm$	0.231 	&	7.681 	$\pm$	-0.535 	\\
Mkn 421	&	39.712 	$\pm$	0.028 	&	41.555 	$\pm$	0.024 	&	46.266 	$\pm$	0.300 	&	42.699 	&	-1.089 	$\pm$	0.137 	&	8.151 	$\pm$	0.292 	&	5.672 	$\pm$	-1.431 	\\
1127-145	&	43.664 	$\pm$	0.036 	&	44.942 	$\pm$	0.031 	&	47.294 	$\pm$	0.300 	&	46.770 	&	-0.563 	$\pm$	0.129 	&	9.164 	$\pm$	0.273 	&	7.729 	$\pm$	-0.914 	\\
4C+29.45	&	43.151 	$\pm$	0.020 	&	44.502 	$\pm$	0.017 	&	46.613 	$\pm$	0.300 	&	45.710 	&	-0.361 	$\pm$	0.143 	&	8.461 	$\pm$	0.258 	&	7.419 	$\pm$	-0.732 	\\
1202-262/PKS 1203-26	&	43.428 	$\pm$	0.036 	&	44.740 	$\pm$	0.031 	&	46.909 	$\pm$	0.300 	&	45.070 	&	-0.189 	$\pm$	0.129 	&	8.719 	$\pm$	0.237 	&	8.001 	$\pm$	-0.572 	\\
1219+285	&	40.336 	$\pm$	0.053 	&	42.090 	$\pm$	0.045 	&	45.814 	$\pm$	0.300 	&	43.250 	&	-0.811 	$\pm$	0.115 	&	7.695 	$\pm$	0.284 	&	5.770 	$\pm$	-1.147 	\\
4C+04.42	&	43.230 	$\pm$	0.036 	&	44.571 	$\pm$	0.031 	&	46.419 	$\pm$	0.300 	&	45.857 	&	-0.303 	$\pm$	0.129 	&	8.257 	$\pm$	0.251 	&	7.325 	$\pm$	-0.673 	\\
4C+21.35	&	42.783 	$\pm$	0.018 	&	44.187 	$\pm$	0.015 	&	46.851 	$\pm$	0.300 	&	46.209 	&	-0.694 	$\pm$	0.145 	&	8.728 	$\pm$	0.280 	&	7.035 	$\pm$	-1.048 	\\
1226+023/3C 273	&	42.983 	$\pm$	0.036 	&	44.359 	$\pm$	0.031 	&	46.624 	$\pm$	0.300 	&	46.540 	&	-0.615 	$\pm$	0.129 	&	8.498 	$\pm$	0.276 	&	6.961 	$\pm$	-0.964 	\\
1253-055/3C 279	&	43.557 	$\pm$	0.036 	&	44.851 	$\pm$	0.031 	&	46.616 	$\pm$	0.300 	&	45.610 	&	-0.166 	$\pm$	0.129 	&	8.420 	$\pm$	0.234 	&	7.742 	$\pm$	-0.553 	\\
1308+326/B2 1308+32	&	43.062 	$\pm$	0.019 	&	44.426 	$\pm$	0.017 	&	46.591 	$\pm$	0.300 	&	46.090 	&	-0.475 	$\pm$	0.144 	&	8.454 	$\pm$	0.267 	&	7.192 	$\pm$	-0.838 	\\
B2 1315+34A	&	43.105 	$\pm$	0.036 	&	44.463 	$\pm$	0.030 	&	47.329 	$\pm$	0.300 	&	46.070 	&	-0.662 	$\pm$	0.130 	&	9.205 	$\pm$	0.278 	&	7.574 	$\pm$	-1.010 	\\
B2 1324+22	&	42.928 	$\pm$	0.036 	&	44.311 	$\pm$	0.031 	&	47.359 	$\pm$	0.300 	&	45.896 	&	-0.709 	$\pm$	0.129 	&	9.237 	$\pm$	0.280 	&	7.513 	$\pm$	-1.055 	\\
1334-127/PKS 1335-127	&	42.782 	$\pm$	0.036 	&	44.187 	$\pm$	0.031 	&	46.094 	$\pm$	0.300 	&	45.180 	&	-0.257 	$\pm$	0.129 	&	7.922 	$\pm$	0.246 	&	7.077 	$\pm$	-0.631 	\\
B3 1343+451	&	43.223 	$\pm$	0.089 	&	44.565 	$\pm$	0.076 	&	47.094 	$\pm$	0.990 	&	46.120 	&	-0.555 	$\pm$	0.084 	&	8.964 	$\pm$	0.971 	&	7.544 	$\pm$	-2.038 	\\
S5 1357+76	&	42.802 	$\pm$	0.064 	&	44.204 	$\pm$	0.055 	&	46.369 	$\pm$	0.210 	&	45.200 	&	-0.331 	$\pm$	0.105 	&	8.212 	$\pm$	0.160 	&	7.227 	$\pm$	-0.351 	\\
B3 1417+385	&	43.197 	$\pm$	0.036 	&	44.542 	$\pm$	0.031 	&	46.734 	$\pm$	0.250 	&	46.104 	&	-0.461 	$\pm$	0.129 	&	8.595 	$\pm$	0.214 	&	7.361 	$\pm$	-0.694 	\\
PKS 1434+235	&	43.078 	$\pm$	0.062 	&	44.440 	$\pm$	0.053 	&	46.489 	$\pm$	0.300 	&	45.775 	&	-0.371 	$\pm$	0.107 	&	8.339 	$\pm$	0.257 	&	7.277 	$\pm$	-0.724 	\\
PKS 1502+106	&	43.488 	$\pm$	0.036 	&	44.792 	$\pm$	0.031 	&	47.221 	$\pm$	0.300 	&	46.297 	&	-0.516 	$\pm$	0.129 	&	9.087 	$\pm$	0.270 	&	7.745 	$\pm$	-0.869 	\\

\bottomrule
\end{tabular}
\begin{flushleft}
Note: The first column is  name; the second column is Logarithm of 151 MHz  radio luminosity; the third column is Logarithm of beam power; the fourth column is the Logarithm of Eddington luminosity; the fifth column is Logarithm of the beaming factor; the sixth column is the Logarithm of the spin function; the seventh column is the Logarithm of black hole spin-mass energy; the eighth column is Logarithm of the black hole irreducible mass.
\end{flushleft}
\end{table*}

\begin{table*}
\centering
\ContinuedFloat 
\captionsetup{labelsep=none} 
\caption{continued}
\begin{tabular}{cccccccc}
\toprule
${\rm [1]}$ & ${\rm [2]}$ & ${\rm [3]}$ & ${\rm [4]}$ & ${\rm [5]}$ & ${\rm [6]}$ & ${\rm [7]}$ & ${\rm [8]}$ \\
 ${\rm name}$ & ${\rm log}L_{\rm {151}}$ & ${\rm log}L_{\rm j}$ & ${\rm log}L_{\rm {Edd}}$ & ${\rm log}L_{\rm {bol}}$ & ${\rm log}{\rm F}$ & ${\rm log}M_{\rm spin}$ & ${\rm log}M_{\rm irr}$ \\
\midrule
1508-055	&	44.229 	$\pm$	0.036 	&	45.426 	$\pm$	0.031 	&	47.084 	$\pm$	0.300 	&	46.520 	&	-0.208 	$\pm$	0.129 	&	8.899 	$\pm$	0.240 	&	8.147 	$\pm$	-0.588 	\\
1510-089/PKS 1510-08	&	42.587 	$\pm$	0.036 	&	44.019 	$\pm$	0.031 	&	46.476 	$\pm$	0.300 	&	45.750 	&	-0.573 	$\pm$	0.129 	&	8.347 	$\pm$	0.273 	&	6.894 	$\pm$	-0.923 	\\
1514-241	&	40.254 	$\pm$	0.036 	&	42.020 	$\pm$	0.031 	&	45.764 	$\pm$	0.300 	&	44.500 	&	-1.100 	$\pm$	0.129 	&	7.649 	$\pm$	0.292 	&	5.146 	$\pm$	-1.439 	\\
B2 1520+31	&	43.048 	$\pm$	0.053 	&	44.415 	$\pm$	0.045 	&	47.034 	$\pm$	0.310 	&	45.900 	&	-0.566 	$\pm$	0.115 	&	8.905 	$\pm$	0.283 	&	7.464 	$\pm$	-0.998 	\\
1546+027/PKS 1546+027	&	41.986 	$\pm$	0.036 	&	43.504 	$\pm$	0.031 	&	46.731 	$\pm$	0.300 	&	45.670 	&	-0.885 	$\pm$	0.129 	&	8.614 	$\pm$	0.287 	&	6.540 	$\pm$	-1.227 	\\
4C+05.64	&	43.754 	$\pm$	0.036 	&	45.020 	$\pm$	0.031 	&	47.294 	$\pm$	0.300 	&	46.057 	&	-0.371 	$\pm$	0.129 	&	9.144 	$\pm$	0.258 	&	8.082 	$\pm$	-0.734 	\\
4C+10.45	&	43.635 	$\pm$	0.036 	&	44.918 	$\pm$	0.031 	&	47.084 	$\pm$	0.300 	&	46.006 	&	-0.351 	$\pm$	0.129 	&	8.931 	$\pm$	0.256 	&	7.907 	$\pm$	-0.716 	\\
1611+343/B2 1611+34	&	43.635 	$\pm$	0.019 	&	44.917 	$\pm$	0.016 	&	47.456 	$\pm$	0.300 	&	46.870 	&	-0.643 	$\pm$	0.144 	&	9.332 	$\pm$	0.278 	&	7.738 	$\pm$	-0.999 	\\
1622-297	&	43.201 	$\pm$	0.036 	&	44.545 	$\pm$	0.031 	&	47.214 	$\pm$	0.300 	&	48.780 	&	-1.171 	$\pm$	0.129 	&	9.099 	$\pm$	0.293 	&	6.456 	$\pm$	-1.509 	\\
1633+382/4C+38.41	&	43.833 	$\pm$	0.023 	&	45.087 	$\pm$	0.020 	&	47.483 	$\pm$	0.300 	&	46.820 	&	-0.555 	$\pm$	0.141 	&	9.353 	$\pm$	0.273 	&	7.933 	$\pm$	-0.912 	\\
4C +47.44	&	43.102 	$\pm$	0.013 	&	44.460 	$\pm$	0.011 	&	46.679 	$\pm$	0.475 	&	45.581 	&	-0.373 	$\pm$	0.149 	&	8.529 	$\pm$	0.438 	&	7.463 	$\pm$	-0.768 	\\
Mkn 501	&	39.879 	$\pm$	0.024 	&	41.698 	$\pm$	0.021 	&	46.951 	$\pm$	0.300 	&	43.204 	&	-1.321 	$\pm$	0.140 	&	8.836 	$\pm$	0.295 	&	5.894 	$\pm$	-1.663 	\\
B3 1708+433	&	42.410 	$\pm$	0.068 	&	43.867 	$\pm$	0.058 	&	46.034 	$\pm$	0.350 	&	45.032 	&	-0.368 	$\pm$	0.102 	&	7.883 	$\pm$	0.308 	&	6.828 	$\pm$	-0.504 	\\
1726+455	&	42.192 	$\pm$	0.036 	&	43.681 	$\pm$	0.031 	&	46.334 	$\pm$	0.300 	&	45.850 	&	-0.722 	$\pm$	0.129 	&	8.212 	$\pm$	0.281 	&	6.463 	$\pm$	-1.068 	\\
1730-130	&	43.602 	$\pm$	0.036 	&	44.890 	$\pm$	0.031 	&	47.414 	$\pm$	0.300 	&	45.830 	&	-0.422 	$\pm$	0.129 	&	9.271 	$\pm$	0.262 	&	8.112 	$\pm$	-0.781 	\\
1739+522	&	43.223 	$\pm$	0.039 	&	44.564 	$\pm$	0.033 	&	47.434 	$\pm$	0.300 	&	46.160 	&	-0.661 	$\pm$	0.127 	&	9.310 	$\pm$	0.278 	&	7.682 	$\pm$	-1.007 	\\
4C+09.57/PKS1749+096	&	42.570 	$\pm$	0.036 	&	44.005 	$\pm$	0.031 	&	46.527 	$\pm$	0.300 	&	44.699 	&	-0.368 	$\pm$	0.129 	&	8.377 	$\pm$	0.258 	&	7.321 	$\pm$	-0.731 	\\
CGRaBS J1800+7828	&	42.784 	$\pm$	0.008 	&	44.188 	$\pm$	0.007 	&	46.374 	$\pm$	0.300 	&	45.851 	&	-0.480 	$\pm$	0.153 	&	8.237 	$\pm$	0.268 	&	6.964 	$\pm$	-0.848 	\\
3C 371	&	40.722 	$\pm$	0.007 	&	42.421 	$\pm$	0.006 	&	46.719 	$\pm$	0.300 	&	43.000 	&	-0.849 	$\pm$	0.155 	&	8.601 	$\pm$	0.286 	&	6.599 	$\pm$	-1.204 	\\
1823+568/4C+56.27	&	43.017 	$\pm$	0.015 	&	44.388 	$\pm$	0.013 	&	47.374 	$\pm$	0.300 	&	44.299 	&	-0.332 	$\pm$	0.148 	&	9.217 	$\pm$	0.255 	&	8.231 	$\pm$	-0.707 	\\
1849+670	&	42.492 	$\pm$	0.036 	&	43.938 	$\pm$	0.031 	&	47.254 	$\pm$	0.300 	&	45.418 	&	-0.763 	$\pm$	0.129 	&	9.134 	$\pm$	0.283 	&	7.303 	$\pm$	-1.108 	\\
1921-293/PKS B1921-293	&	42.273 	$\pm$	0.036 	&	43.751 	$\pm$	0.031 	&	46.809 	$\pm$	0.300 	&	45.020 	&	-0.645 	$\pm$	0.129 	&	8.684 	$\pm$	0.277 	&	7.088 	$\pm$	-0.992 	\\
2145+067/4C+06.69	&	43.409 	$\pm$	0.036 	&	44.724 	$\pm$	0.031 	&	46.984 	$\pm$	0.300 	&	46.766 	&	-0.583 	$\pm$	0.129 	&	8.856 	$\pm$	0.274 	&	7.381 	$\pm$	-0.933 	\\
2155-152/PKS 2155-152	&	42.948 	$\pm$	0.036 	&	44.329 	$\pm$	0.031 	&	45.704 	$\pm$	0.300 	&	44.682 	&	0.032 	$\pm$	0.129 	&	7.423 	$\pm$	0.203 	&	7.095 	$\pm$	-0.398 	\\
BL Lac	&	40.470 	$\pm$	0.036 	&	42.204 	$\pm$	0.031 	&	46.324 	$\pm$	0.300 	&	43.519 	&	-0.957 	$\pm$	0.129 	&	8.207 	$\pm$	0.289 	&	5.992 	$\pm$	-1.297 	\\
PKS 2209+236	&	42.615 	$\pm$	0.036 	&	44.044 	$\pm$	0.031 	&	46.574 	$\pm$	0.450 	&	45.786 	&	-0.596 	$\pm$	0.129 	&	8.446 	$\pm$	0.427 	&	6.947 	$\pm$	-1.458 	\\
2223-052/3C 446	&	44.516 	$\pm$	0.036 	&	45.672 	$\pm$	0.031 	&	46.531 	$\pm$	0.300 	&	46.601 	&	0.056 	$\pm$	0.129 	&	8.237 	$\pm$	0.199 	&	7.948 	$\pm$	-0.381 	\\
PKS 2227-08	&	43.788 	$\pm$	0.036 	&	45.049 	$\pm$	0.031 	&	46.899 	$\pm$	0.300 	&	46.664 	&	-0.375 	$\pm$	0.129 	&	8.749 	$\pm$	0.258 	&	7.681 	$\pm$	-0.737 	\\
2230+114/CTA 102	&	43.668 	$\pm$	0.036 	&	44.946 	$\pm$	0.031 	&	46.894 	$\pm$	0.300 	&	46.870 	&	-0.469 	$\pm$	0.129 	&	8.756 	$\pm$	0.266 	&	7.505 	$\pm$	-0.825 	\\
2251+158/3C 454.3	&	43.860 	$\pm$	0.036 	&	45.110 	$\pm$	0.031 	&	46.939 	$\pm$	0.300 	&	46.650 	&	-0.352 	$\pm$	0.129 	&	8.786 	$\pm$	0.256 	&	7.761 	$\pm$	-0.717 	\\
TXS 2331+073	&	41.628 	$\pm$	0.036 	&	43.198 	$\pm$	0.031 	&	46.484 	$\pm$	0.300 	&	45.932 	&	-1.025 	$\pm$	0.129 	&	8.368 	$\pm$	0.490 	&	6.017 	$\pm$	-1.164 	\\
2345-16/PKS 2345-16	&	42.811 	$\pm$	0.036 	&	44.211 	$\pm$	0.031 	&	46.709 	$\pm$	0.300 	&	45.359 	&	-0.459 	$\pm$	0.129 	&	8.570 	$\pm$	0.265 	&	7.339 	$\pm$	-0.815 	\\
\bottomrule
\end{tabular}
\begin{flushleft}
Note: The first column is  name; the second column is Logarithm of 151 MHz  radio luminosity; the third column is Logarithm of beam power; the fourth column is the Logarithm of Eddington luminosity; the fifth column is Logarithm of the beaming factor; the sixth column is the Logarithm of the spin function; the seventh column is the Logarithm of black hole spin-mass energy; the eighth column is Logarithm of the black hole irreducible mass.
\end{flushleft}
\end{table*}

\section{The RESULTS}
\subsection{fundamental plane of blazars}
\begin{figure*}
  \centering
  \includegraphics[width=0.49\textwidth]{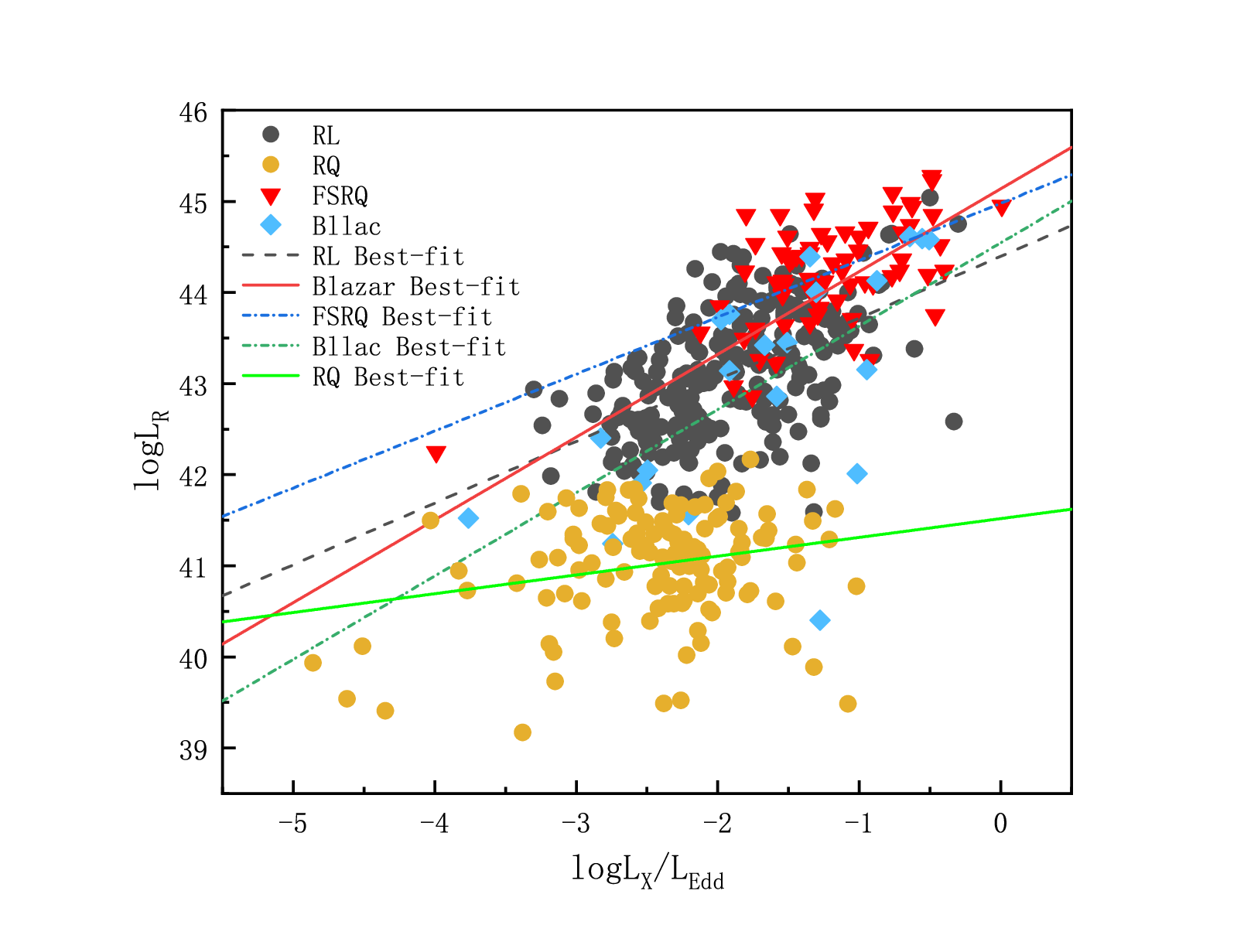}
  \includegraphics[width=0.49\textwidth]{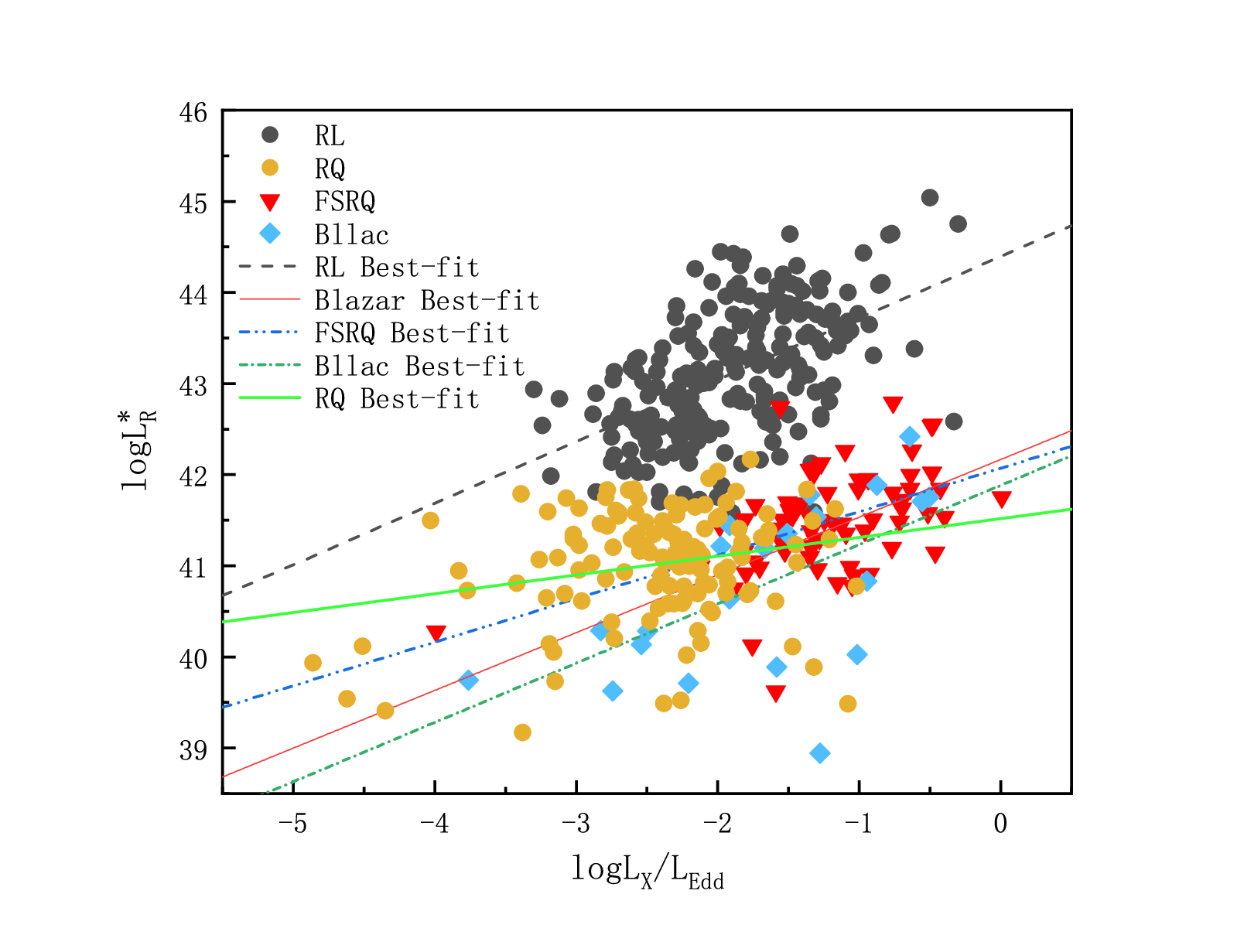}
  \caption{Left panel: The correlation between $L_{\rm{X}}$/$L_{\rm{Edd}}$ and $L_{\rm{R}}$. Right panel: The correlation between $L_{\rm{X}}$/$L_{\rm{Edd}}$ and $L_{\rm{R}}$ with beaming corrected.}
\label{fig:2}
\end{figure*}

To examine the fundamental plane of Blazars, we should first explore the correlation between radio luminosity ($L_{\rm{R}}$), X-ray luminosity ($L_{\rm{X}}$), and black hole mass (M) which also can be converted to Eddington luminosity $L_{\rm{Edd}}$. We use the method of \cite{Merloni2003} to determine the exist of fundamental plane. The results of the best fit is showed in Table \ref{table:3}. We found that the fundamental plane exists for Blazars, just as previous research has shown for AGN \citep{Merloni2003,Unal2020,Bariuan2022}. This means that there is a correlation between $L_{\rm{X}}$/$L_{\rm{Edd}}$ and $L_{\rm{R}}$ \citep{Merloni2003,Zhang2018}. And we found the best fit of blazars is very similar to radio-loud quasars (RL). The data points are distributed along the direction of the extended line of fit. This confirms the current theoretical classification of Blazars as a subtype of radio-loud active galactic nuclei. After correcting for the beam effect, the radio data points of FSRQs change from near the radio loud quasar region ($logL_{\rm{R}}$=42.5-45) with average value 45.26 to near the radio quiet quasar region after correcting for the beam effect ($logL_{\rm{R^{*}}}$=39.5-43) with average value 42.65. It suggests that the differences for Blazars between left plane and right plane of Figure \ref{fig:2} may be caused by the beaming effect induced by relativistic jets. This further underscores the importance of accounting for beaming effects when conducting fundamental plane analysis of Blazars.

\begin{table}
\caption{The correlation analysis for $L_{R}$ and $\frac{L_{\rm{X}}}{L_{\rm{Edd}}}$}
\label{table:3}
\begin{tabular}{lll}
\hline
Type                          &       Best fit       &      R \\
\hline
RL          &       log$L_{R}$=${0.677}\pm{0.074}$ log($\frac{L_{\rm{X}}}{L_{\rm{Edd}}}$)+${44.396}\pm {0.147}$               &      0.272      \\
RQ          &      log$L_{R}$=${0.206}\pm{0.077}$ log($\frac{L_{\rm{X}}}{L_{\rm{Edd}}}$)+${41.518}\pm {0.193}$                &    0.054       \\
\hline
FSRQ        &      log$L_{R}$=${0.625}\pm{0.103}$ log($\frac{L_{\rm{X}}}{L_{\rm{Edd}}}$)+${44.977}\pm {0.139}$                &     0.355      \\
Bllac       &        log$L_{R}$=${0.909}\pm{0.115}$ log($\frac{L_{\rm{X}}}{L_{\rm{Edd}}}$)+${45.138}\pm {0.172}$              &    0.391       \\
Blazar      &          log$L_{R}$=${0.909}\pm{0.115}$ log($\frac{L_{\rm{X}}}{L_{\rm{Edd}}}$)+${45.138}\pm {0.172}$            &      0.416     \\
\hline
${\rm {FSRQ}}^*$    &     log$L_{R}$=${0.478}\pm{0.107}$ log($\frac{L_{\rm{X}}}{L_{\rm{Edd}}}$)+${42.071}\pm {0.145}$          &     0.229      \\
${\rm {BL Lac}}^*$   &     log$L_{R}$=${0.651}\pm{0.205}$ log($\frac{L_{\rm{X}}}{L_{\rm{Edd}}}$)+${41.883}\pm {0.386}$          &      0.347     \\
${\rm {Blazar}}^*$  &     log$L_{R}$=${0.634}\pm{0.095}$ log($\frac{L_{\rm{X}}}{L_{\rm{Edd}}}$)+${42.166}\pm {0.142}$         &      0.338      \\
\hline
\end{tabular}

Note: The R is R-Square as correlation coefficient; p is significance level; * indicates that the corresponding data has been corrected with beaming factor.
\end{table}

\begin{figure*}
  \centering
  \includegraphics[width=0.49\textwidth]{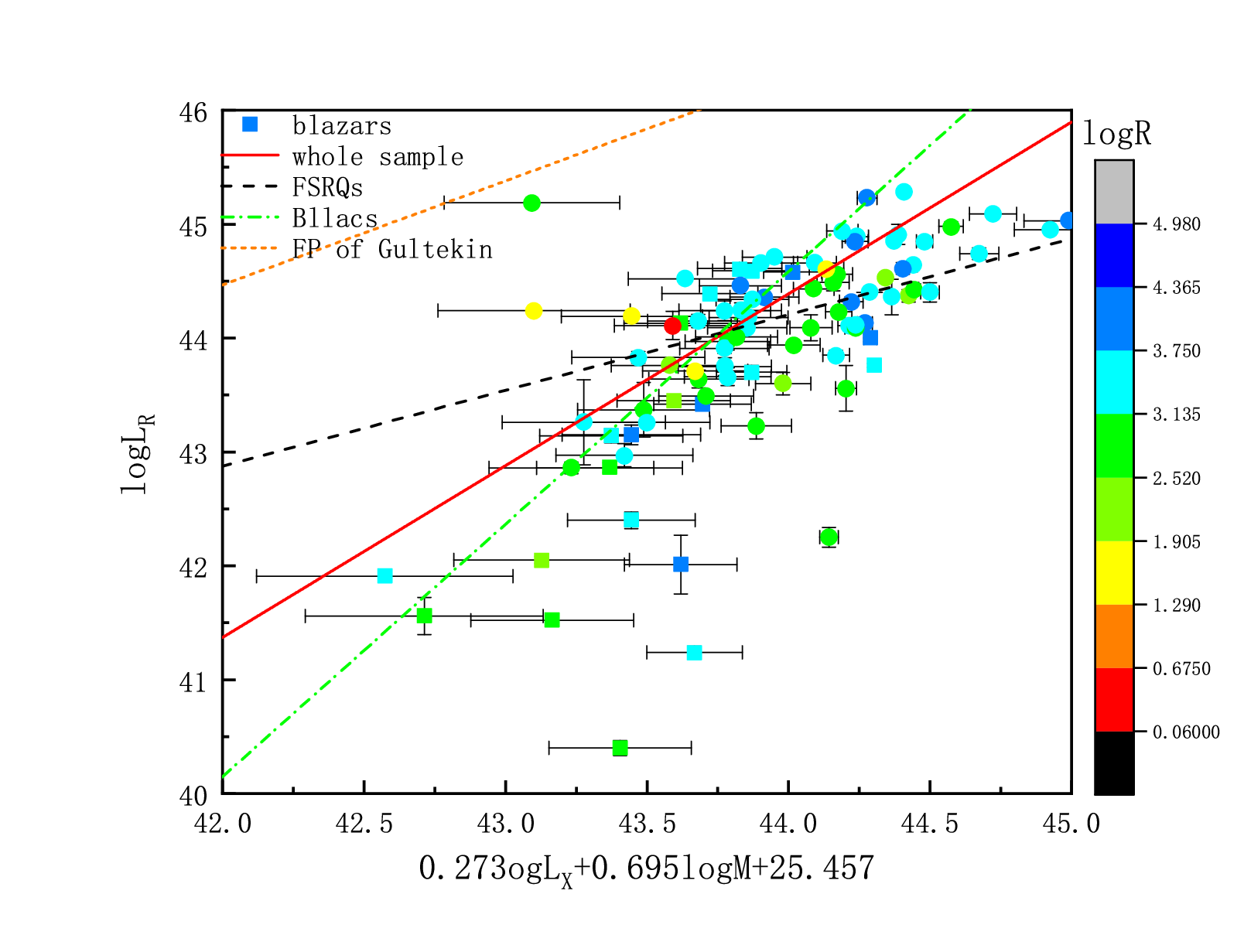}
  \includegraphics[width=0.49\textwidth]{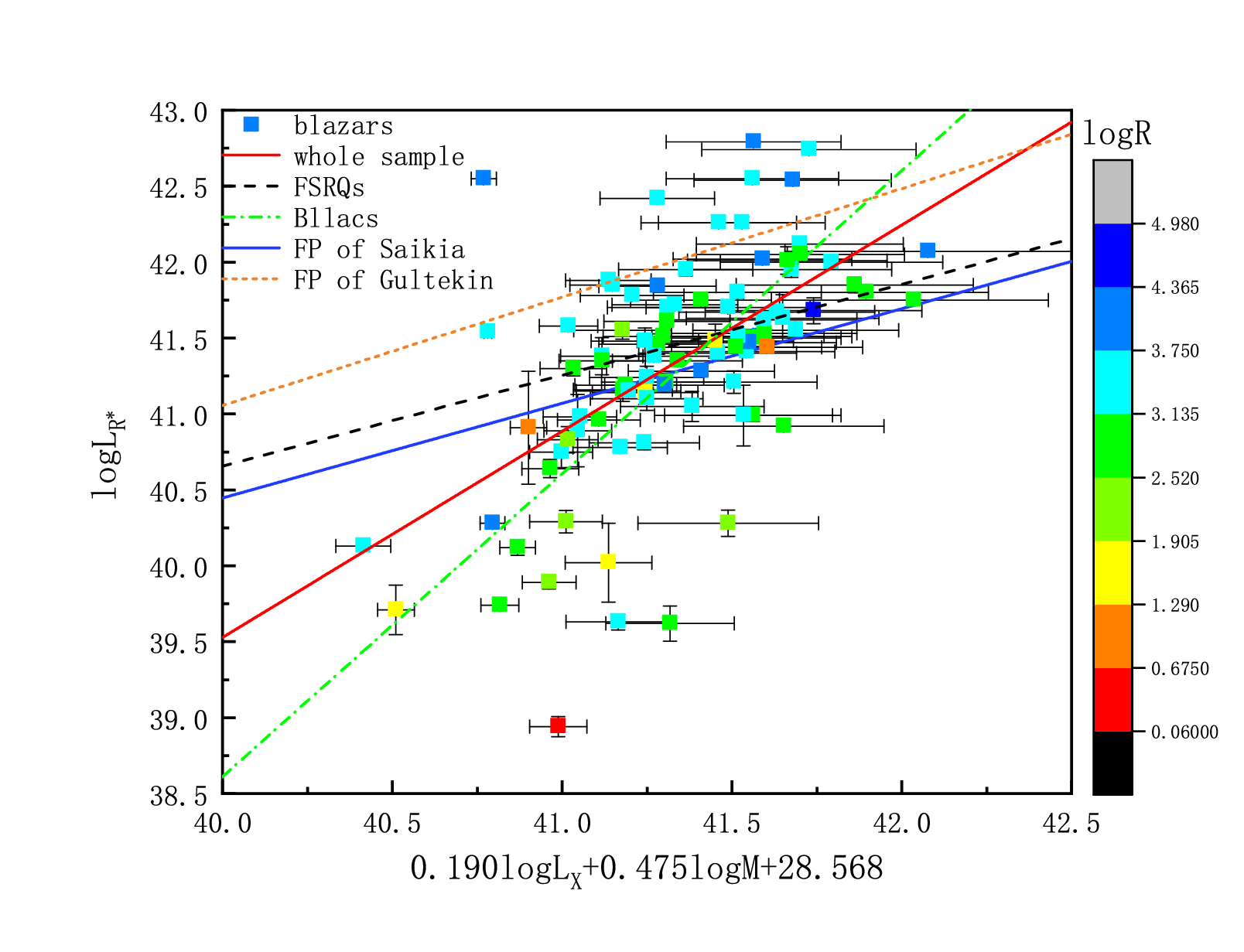}
  \caption{Left panel: The fundamental plane of Blazars. Right panel: The fundamental plane of Blazars with beaming corrected. R is the radio loudness. FP of Saikia is from Saikia et al. 2018, FP of Gultekin is from Gultekin et al. 2019. }
\label{fig:3}
\end{figure*}

In this paper, the fundamental plane of Blazars we defined is consistent with \cite{Merloni2003}. We use a multiple linear regression equation to obtain the fundamental relationship and verify its reliability.
\begin{eqnarray}
{\rm log}L_{\rm R}=a{\rm log}L_{\rm X}+b{\rm log}M+c,
\label{Eq8}
\end{eqnarray}
where $L_{\rm R}$ is 5 GHz luminosity, $L_{\rm X}$ is the X-ray luminosity in 2-10 keV, M is the BH mass of Blazars. $a$ and $b$ and $c$ are fitting parameters \citep{Zhang2018}. Multiple linear regression analysis was performed to obtain the best fitting linear regression coefficient \citep{Merloni2003,Kording2006a,Zhang2018,Zhang2023a}. Linear regression analysis enables the derivation of three variables, namely a, b, and c, each with associated errors \citep{Wang2017}. The fundamental plane for Blazars, log$L_{R}$=${0.273}_{+0.059}^{-0.059}$log$L_X$+${0.695}_{+0.191}^{-0.191}$log$M$+${25.457}_{+2.728}^{-2.728}$ with R-Square=0.355, is showed in left plane of Figure \ref{fig:3}. We examined the fundamental plane relation for Blazars. It shows that the jet energy of Blazars has a correlation with accretion and spin of BH. It is consistent with the theory of BZ model \citep{Blandford1977}. Due to the strong relativistic jets of Blazars, 5 GHz radio luminosity originating from jet can be affected by the beaming effect. It may influence the analysis results of the fundamental plane. Therefore, we measured the intrinsic 5 GHz radio luminosity for conducting the fundamental plane analysis of Blazars. The fundamental plane of Blazars after correcting anisotropism, log$L_{R}$=${0.190}_{+0.049}^{-0.049}$log$L_X$+${0.475}_{+0.157}^{-0.157}$log$M$+${28.568}_{+2.245}^{-2.245}$ with R-Square=0.279, is showed in right plane of Figure \ref{fig:3}. The existing FP relations from previous work \citep{Saikia2018,Gultekin2019} are also included in Figure 3 for visualizing the results. The difference between our result and previous work has decreased after correcting anisotropism. This suggests that the fundamental plane of FSRQs are more similar to the fundamental plane of AGNs \citep{Saikia2018} with beaming corrected. Both fundamental plane relations of Blazars are positive correlation and generally similar. However, there is a slight decrease in the R-Square of the fundamental plane, comparing to the original version, indicating that the beaming effect can have some impact on the fundamental plane analysis. The beaming effect enhances the role of jets in the fundamental plane relationship. The analysis results are showed in Table \ref{table:4}. The R-values (0.595 and 0.528) are above 0.5 and the significance of each variable is below the threshold of 0.05. The fitting results indicate that the correlation of fundamental plane for Blazars is strong. It supports the validity of the fundamental relationship. We also separately fitted FSRQs and BL Lacs by the same fundamental plane which is calculated by total sample. It can be observed that there are differences between the best-fitting lines of FSRQs and BL Lacs, and the best fit for the overall sample lies between these two. Furthermore, this characteristic remains even after the elimination of the beaming effect. The fundamental plane relationship still exists when Blazars and BL Lacs are discussed separately.

\begin{table}
\centering
\caption{The correlation analysis for the fundamental plane of Blazars}
\label{table:4}
\begin{tabular}{lllll}
\hline
 Type                                    & R-Square                      & p($L_{\rm{X}}$)              & p(M)                       &      DW         \\
\hline
 Blazars                                  & 0.355                         & 0.001                        & 0.001                      &     2.011       \\
 $\rm{{Blazars}^*}$                       & 0.279                         &  0.001                       & 0.003                      &     2.024       \\
\hline
\end{tabular}
\begin{flushleft}
Note:R-Square is correlation coefficient; p($L_{\rm{X}}$) is significance level of log$L_{\rm{X}}$; p(M) is significance level of logM ; * indicates that the corresponding data has been corrected with beaming factor. DW is the Durbin-Watson statistic which is a statistical method used to assess the presence of autocorrelation in time series data.
\end{flushleft}
\end{table}

We also used redshift as third variable to conduct a partial correlation analysis of the correlation and discussed the effect of distance on result. The R-square value decreases from 0.279 to 0.232 and the significance level of both variables increased to 0.029 (still less than 0.05) after excluding the effect of redshift (distance) on the fundamental plane which has been corrected with beaming factor. It indicates that the result of fundamental plane is partly effected by distance. The correlation of fundamental plane still exists after removing the effect of distance from both the variables.

\subsection{fundamental plane of blazars base on BH spin energy}
\begin{figure*}
  \centering
  \includegraphics[width=0.49\textwidth]{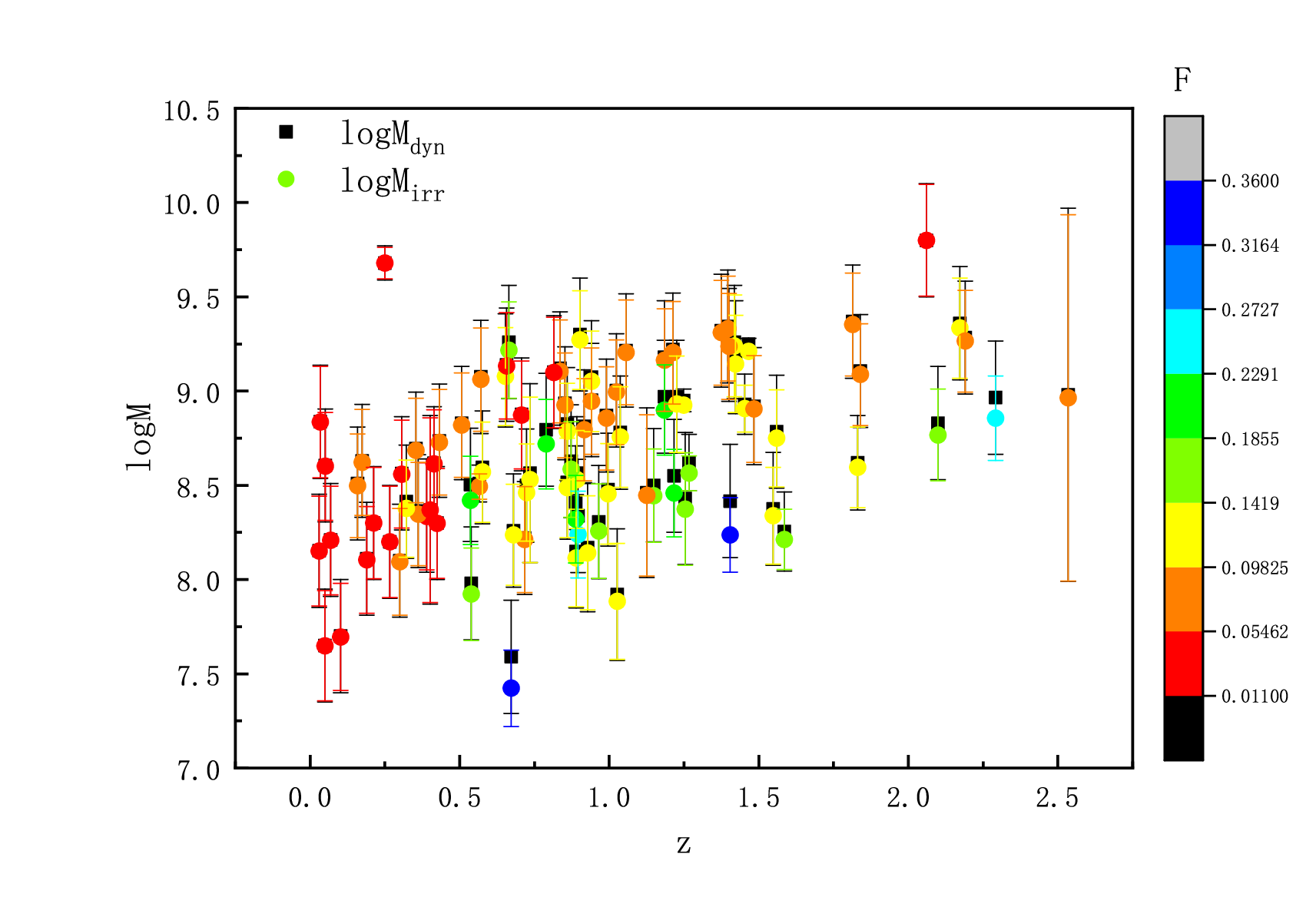}
  \includegraphics[width=0.49\textwidth]{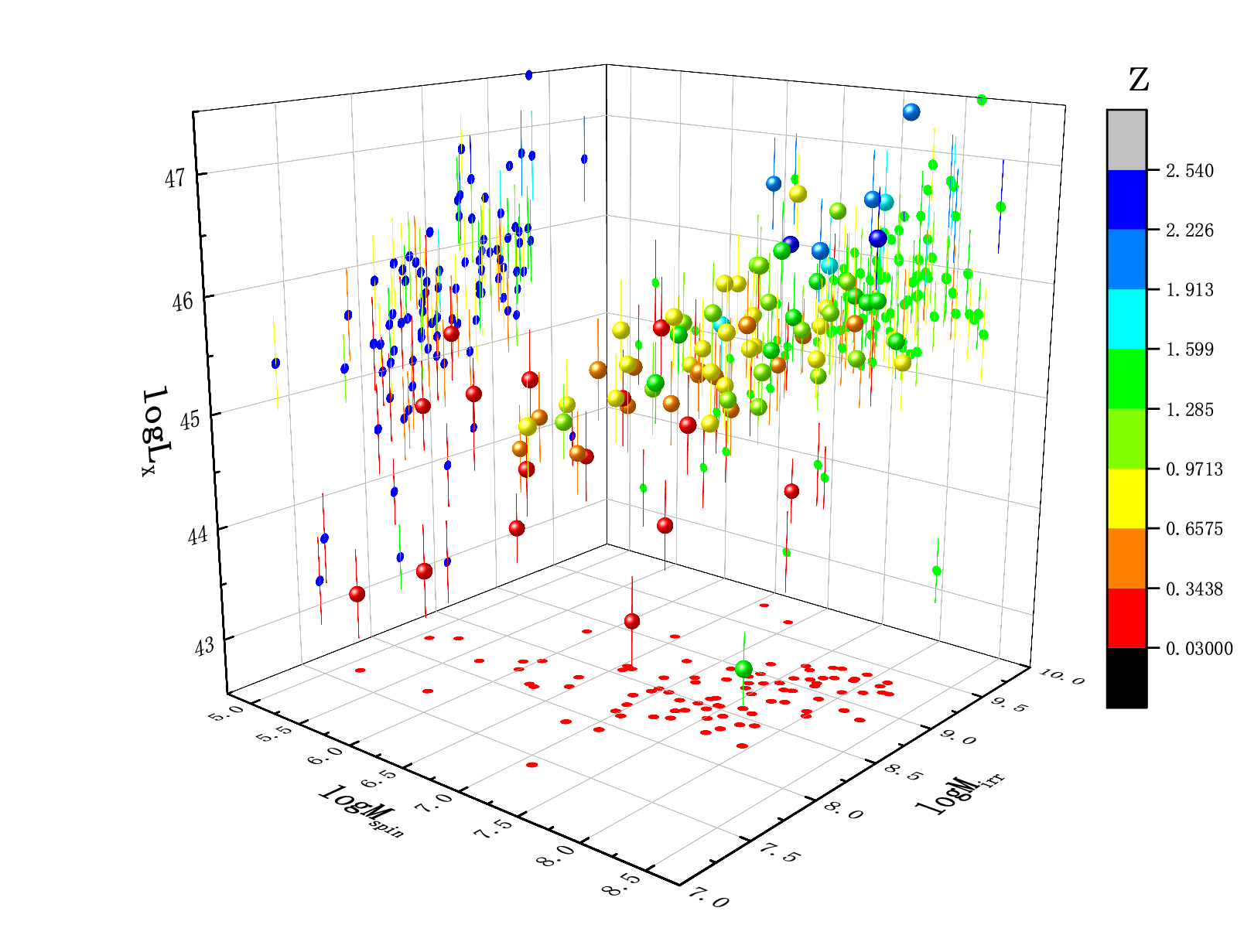}
  \includegraphics[width=0.49\textwidth]{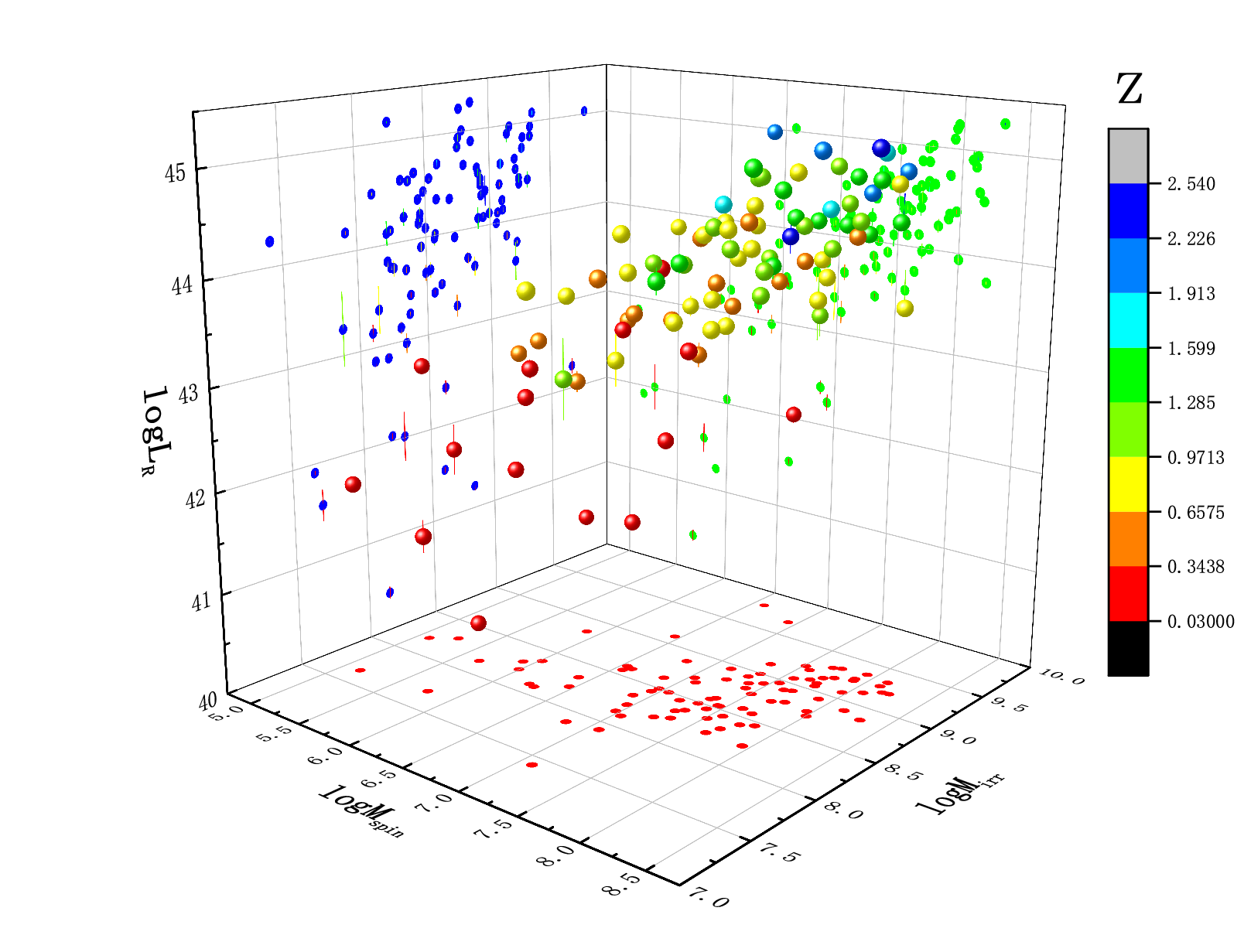}
  \includegraphics[width=0.49\textwidth]{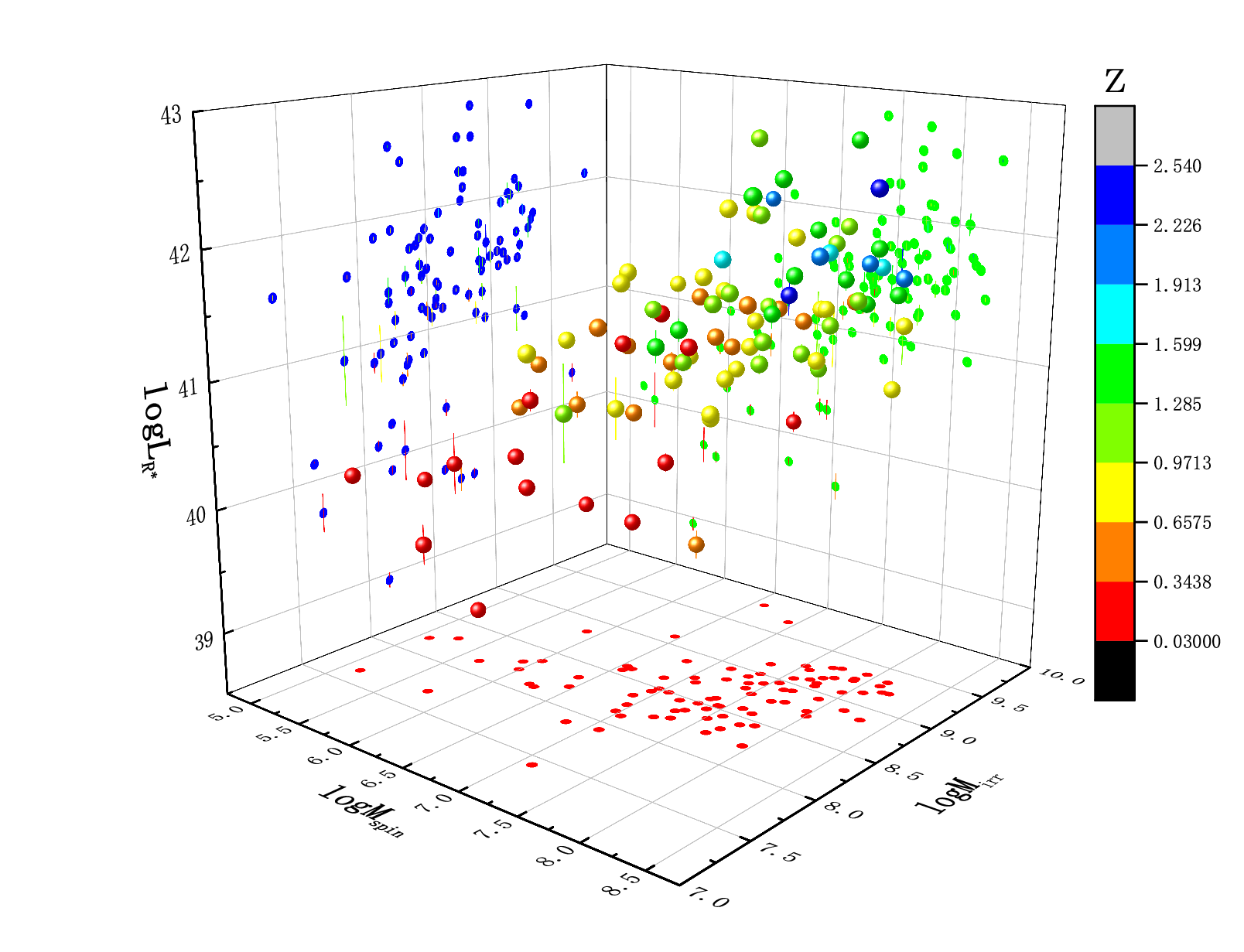}
  \caption{ Upper left panel: The differences between ${\rm log}M_{\rm spin}$  and ${\rm log}M_{\rm irr}$.  Upper right panel: The correlation between ${\rm log}L_{\rm X}$, ${\rm log}M_{\rm spin}$ and ${\rm log}M_{\rm irr}$. Lower left panel: The correlation between ${\rm log}L_{\rm R}$, ${\rm log}M_{\rm spin}$ and ${\rm log}M_{\rm irr}$.  Lower right panel: The correlation between ${\rm log}L_{\rm R^*}$, ${\rm log}M_{\rm spin}$ and ${\rm log}M_{\rm irr}$ }
\label{fig:4}
\end{figure*}

Before using BH spin energy as a variable for fundamental analysis, we need to discuss its effect on the black hole mass and its correlation with other variables. The relation between z and $M_{\rm dyn}$, $M_{\rm irr}$ is presented in upper left panel of Figure \ref{fig:4}. The differences between  $M_{\rm dyn}$ and $M_{\rm irr}$ are evident. This is due to the contribution of $M_{\rm spin}$ (where M$\equiv$$M_{\rm dyn}$=$M_{\rm irr}$+$E_{\rm spin}$$c^{-2}$=$M_{\rm irr}$+$M_{\rm spin}$), where the differences in mass are determined by the numerical value of the black hole spin. While the difference is negligible in samples with low spin, samples characterized by high spin exhibit a more significant discrepancy \citep{Daly2022,Zhang2023a}. The 3D correlation diagram: upper right panel, lower left panel and lower right panel in Figure \ref{fig:4}, show that there are obvious correlations among ${\rm log}L_{\rm X}$, ${\rm log}L_{\rm R}$ , ${\rm log}M_{\rm spin}$ and ${\rm log}M_{\rm irr}$. They are positively correlated with each other. The results of R-Square are listed in Table \ref{table:5}. This correlations existed after the effects of beam factors were taken into account. The analysis results indicate that the correlation observed in the fundamental plane relation of $M_{\rm dyn}$ is partly contributed by the $M_{\rm spin}$. The analysis results also indicate that the correlation of black hole spin is more strongly correlated with the fundamental plane relation than irreducible black hole mass. It is important to analyze the role of black hole spin energy in the fundamental plane relation.

\begin{table}
\centering
\caption{Correlation coefficient between $L_{\rm X}$, $L_{\rm R}$ and $M_{\rm irr}$, $M_{\rm spin}$}
\label{table:5}
\begin{tabular}{lllll}
\hline
 Type                      & ${\rm log}L_{\rm X}$                        & ${\rm log}L_{\rm R}$      &      ${\rm log}L_{\rm R^*}$       \\

 & R-Square & R-Square & R-Square \\
\hline
 $M_{\rm irr}$                           &     0.211                                   &      0.158                 &            0.121                        \\
 $M_{\rm spin}$.                         &     0.362                                   &      0.604                 &            0.501                        \\
\hline
\end{tabular}
\end{table}

\begin{figure*}
  \centering
  \includegraphics[width=0.49\textwidth]{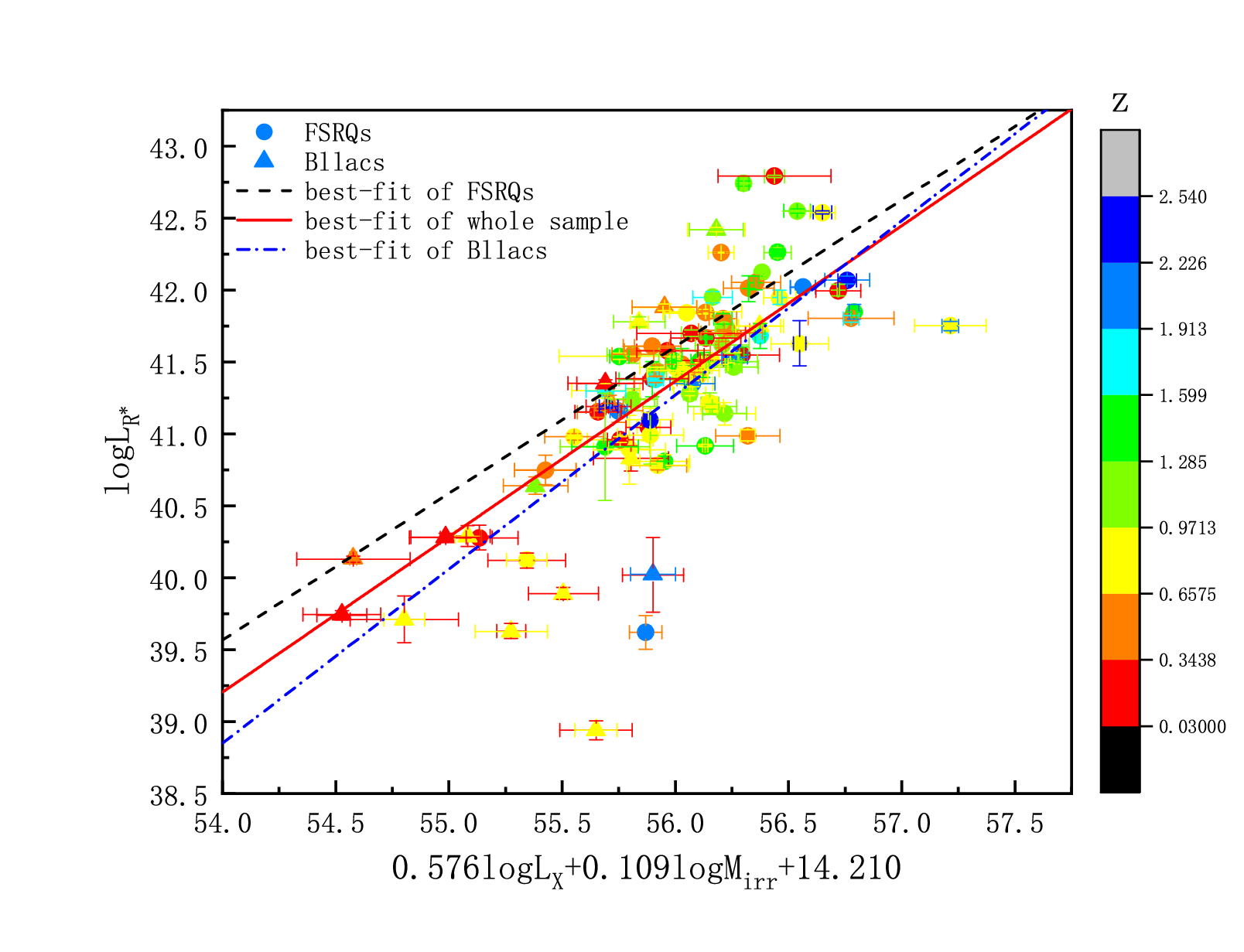}
  \includegraphics[width=0.49\textwidth]{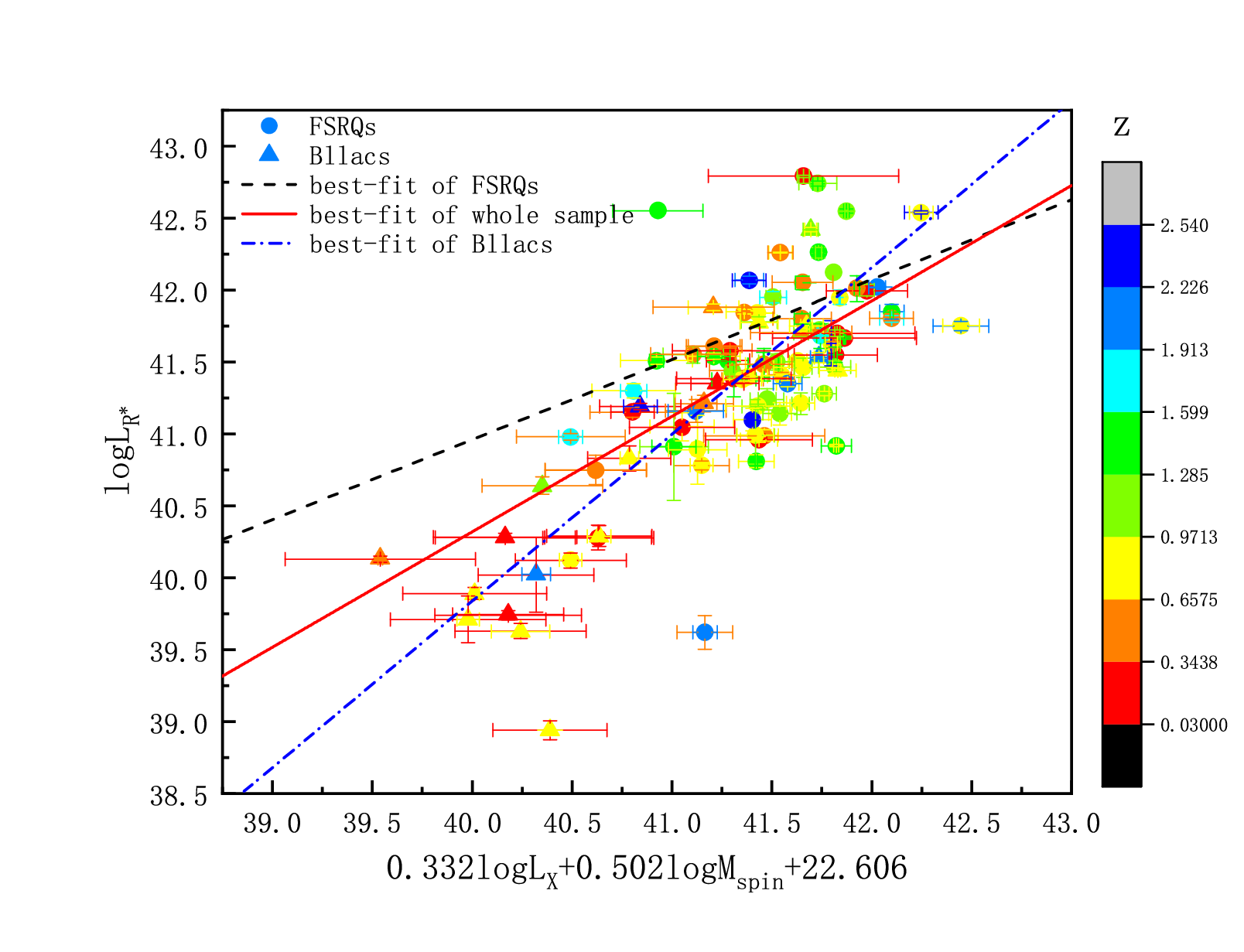}
  \caption{Left panel: The fundamental plane of Blazars based on ${\rm log}M_{\rm irr}$ after beam-correction. Right panel: The fundamental plane of Blazars based on ${\rm log}M_{\rm spin}$ after beam-correction.}
\label{fig:5}
\end{figure*}

Consistent with \cite{Merloni2003}, we examined the fundamental plane of Blazars based on $M_{\rm spin}$. We replace the $M_{\rm dyn}$ variable with $M_{\rm irr}$ and $M_{\rm spin}$, and discuss them separately in this paper. The results for new fundamental plane is showed in Table \ref{table:6}. After separately analyzing $M_{\rm spin}$ and $M_{\rm irr}$, as shown in Figure \ref{fig:5}. All the correlations are positive, and there are differences between the fundamental plane of $M_{\rm irr}$ and $M_{\rm spin}$ for Blazars. This may indicate that both $M_{\rm irr}$ and $M_{\rm spin}$ contribute to the correlation foundation for fundamental plane of Blazars. According to the results for new fundamental plane, $M_{\rm spin}$ contributes more correlation than $M_{\rm irr}$. It means that the effect of BH spin on fundamental plane is relatively large. Since XRBs applie to fundamental relationships and its jet mechanism is also similar to AGNs. This effect is similar to the radio loud and radio quiet sources in the $L_{r}-L_{x}$ correlation of XRBs \citep{Fender2003,Gallo2003} for the sources of XRBs which have black hole spins. Therefore, it is necessary to discuss the effect of $M_{\rm spin}$ on the traditional fundamental plane and consider the $M_{\rm spin}$ as a important factor for the study of fundamental plane.

\begin{table}
\centering
\caption{The correlation analysis for the fundamental plane based on $M_{\rm{irr}}$ and $M_{\rm{spin}}$}
\label{table:6}
\begin{tabular}{lllll}
\hline
 Type(left plane)                        & R-Square                   & p($L_{\rm{X}}$)          & p($M_{\rm{irr}}$)  &  DW \\
\hline
 $\rm{{Blazars}^*}$                             & 0.429 & 0.001 & 0.010 & 1.9 \\
\hline
 Type(right plane)                        & R-Square                   & p($L_{\rm{X}}$)          & p($M_{\rm{spin}}$)    &  DW  \\
\hline
 $\rm{{Blazars}^*}$                             & 0.575                      &  0.001   & 0.001     & 1.7 \\
\hline
\end{tabular}
\begin{flushleft}
Note: R-Square is correlation coefficient;  p($L_{\rm{X}}$) is significance of log$L_{\rm{X}}$; p($M_{\rm{irr}}$) is significance of log$M_{\rm{irr}}$ ; p($M_{\rm{spin}}$) is significance level of log$M_{\rm{spin}}$ ; * indicates that the corresponding data has been corrected with beaming factor. DW is the Durbin-Watson statistic which is a statistical method used to assess the presence of autocorrelation in time series data.
\end{flushleft}
\end{table}

\section{CONCLUSIONS}
In this paper, we repeated the fundamental plane of previous work \citep{Merloni2003} with Blazars. A total of 91 samples were obtained through sample collection. We examined a traditional fundamental plane of black hole activity for Blazars. The fundamental plane is log$L_{R}$=${0.273}_{+0.059}^{-0.059}$log$L_X$+${0.695}_{+0.191}^{-0.191}$log$M$+${25.457}_{+2.728}^{-2.728}$ with R-Square=0.355. Although this fundamental plane is similar to the previous work of AGN in general \citep{Merloni2003,Unal2020,Bariuan2022}, there are quite differences in the coefficients. The fundamental plane for our Blazars after considering the effect of beam factor is log$L_{R}$=${0.190}_{+0.049}^{-0.049}$log$L_X$+${0.475}_{+0.157}^{-0.157}$log$M$+${28.568}_{+2.245}^{-2.245}$ with R-Square=0.279 . The fundamental plane of Blazars with beaming effect corrected is more similar to the radio-quiet quasars of previous work \citep{Bariuan2022}. This could be attributed to the jet opening angles of Blazars, which results in a significant influence of beaming effect on the jet emission. The work of \cite{Bariuan2022} has indicated that radio loudness has an impact on the fundamental plane  within quasar samples. To accurately model the dichotomy between radio-loud and radio-quiet quasars, two different fundamental plane models are required. It has been previously concluded that radio-loud and radio-quiet quasars may be governed by distinct fundamental plane models \citep{Bariuan2022}. Typically, the emission mechanism for radio-quiet samples is dominated by accretion flows, while radio-loud samples are likely a combination of jet and disk. However, in this study, as shown in Figure \ref{fig:2}, our Blazars tend to exhibit characteristics of radio-loud samples before correcting anisotropism, while their characteristics lean towards radio-quiet samples after considering the effect of beam factor. We speculate that the difference in jet opening angles leads to the manifestation of both jet-dominated and accretion disk-dominated fundamental plane models for Blazars. Due to the fact that all samples are radio-loud objects, and the distribution of radio loudness data points in Figure \ref{fig:3} wasn't affected by the beaming effect. We found that Blazars are more suitable for the jet-dominated fundamental plane models.

Standard fundamental models can accurately describe the production of black hole jets \citep{Unal2020}, but they still have some deviations. In this paper, we consider the influence of black hole spin to reduce these deviations in the fundamental models. As the spin increases, a larger portion of the gravitational energy can be converted into radiation due to the closer proximity of accreted matter to the event horizon. The correlation indicates a positive relationship between the jet power and the black hole spin-mass energy, which is validated by correlation analysis. Therefore, the effect of black hole spin should be considered for the study of fundamental plane. To improve accuracy of conclusion, our goal is to include as many variables as possible. Thus, we introduce $M_{\rm spin}$ as a variable in the fundamental model, which is calculated by the spin and mass of BH. The new fundamental plane in this paper  (log$L_{\rm R}$=${0.332}_{+0.081}^{-0.081}$log$L_{\rm X}$+${0.502}_{+0.091}^{-0.091}$log$M_{\rm spin}$+${22.606}_{+3.346}^{-3.346}$ with R-Square=0.575) shows that the fundamental plane based on $M_{\rm spin}$ has a stronger relation compared to the fundamental plane based on the mass of BH also know as $M_{\rm dyn}$ (log$L_{\rm R}$=${0.190}_{+0.049}^{-0.049}$log$L_{\rm X}$+${0.475}_{+0.157}^{-0.157}$log$M$+${28.568}_{+2.245}^{-2.245}$ with R-Square=0.279). Hence, the parameter $M_{\rm spin}$ holds significant importance in the fundamental plane of Blazars \citep{Zhang2023a}. Our findings underscore the crucial role of black hole spin in the accretion and outflow processes \citep{Unal2020}. We confirmed the correlation among jet, accretion, and spin of BH which indicates the applicability of BZ model for Blazars. It is consistent with the conclusions of previous work \citep{Blandford1977, Blandford2019, Unal2020} and could provide a observational indication for the existence of the Blandford-Znajek process.

Due to the flux variation of Blazars, the observed value of the sample will change with time, especially for radio luminosity, if the observation time interval is large, it may produce errors, which will affect the fundamental plane analysis’ results. This change may cause the correlation of R-Square to decrease. Agns and blazars have different bands of radiation that are correlated with each other and there is a significant time lag between these radiations \citep{Xie2017}. In order to improve the accuracy of the analysis results, we limited the observation time of the collected data sources in 2-5 years. However, the time interval between radio and X-ray emissions in Blazars generally ranges from months to years. AGNs exhibit variability, with the potential for amplitude reaching approximately 100$\%$ on timescales that are shorter than the observation time interval \citep{Ho2001}. This uncertainty may govern the actual observed uncertainty of flux \citep{Xie2017}. Therefore, to investigate the intrinsic and physically related Fundamental Plane, it is important to account for the radio/X-ray time interval. But implementing such a correction for Blazars is exceedingly difficult due to the requirement of long-term, coordinated monitoring of individual sources in both radio and X-ray bands \citep{Xie2017}. Additionally, this monitoring must capture at least one outburst in each Blazars to obtain a dependable timelag measurement. Consequently, only a limited number of AGNs have undergone such intensive monitoring to date, as it poses significant challenges \citep{Bell2011,King2013}. In this paper the uncertainty because of non-simultaneous (radio and X-ray) observations we taken is from \cite{Ho2001} which adopted a slightly smaller scatter in radio luminosities (0.2 dex).

\section*{Acknowledgments}
We gratefully thank the anonymous referee for the very helpful comments, which helped us to greatly improve this paper. This work is partially supported by Natural Science Foundation of Yunnan(202101AU070006), National Natural Science Foundation of China (12063006), Young and middle-aged academic and technical reserve talents of Yunnan Province (202205AC160087) and Yunnan Provincial Department of Education Research Fund (2021J0671). The authors gratefully acknowledge the computing support provided by the JRT Science Data Center at Yuxi Normal University. QLH gratefully acknowledges the financial support from the Hundred Talents Program of Yuxi (grant 2019).

\section*{Data availability}
The data underlying this article are available in the article and in its online supplementary material.


\clearpage

\begin{thebibliography}{}
\bibitem[Abdo et al.(2010)]{Abdo2010}Abdo, A. A., Ackermann, M., Ajello, M., et al. 2010, ApJS, 188, 405. \url{https://doi.org/10.1088/0067-0049/188/2/405}

\bibitem[Ackermann et al.(2011)]{Ackermann2011}Ackermann, M., Ajello, M., Albert, A., et al. 2011, ApJ, 743, 171. \url{https://doi.org/10.1088/0004-637X/743/2/171}

\bibitem[Bariuan et al.(2022)]{Bariuan2022}Bariuan, L. G., Snios, B.,Sobolewska, M., et al. 2022, MNRAS, 513, 4673. \url{https://doi.org/10.1093/mnras/stac1153}

\bibitem[Begelman et al.(1984)]{Begelman1984}Begelman, M. C., Blandford, R. D., Rees, M, J., 1984, Rev. Mod. Phys, 56, 255. \url{https://doi.org/10.1103/RevModPhys.56.255}

\bibitem[Bell et al.(2011)]{Bell2011}Bell, M. E., Tzioumis, T., Uttley, P., et al. 2011, MNRAS, 411, 402. \url{https://doi.org/10.1111/j.1365-2966.2010.17692.x}

\bibitem[Blandford et al.(1977)]{Blandford1977}Blandford, R. D., Znajek, R. L., 1977, MNRAS, 179, 433.  \url{https://doi.org/10.1093/mnras/179.3.433}

\bibitem[Blandford et al.(1978)]{Blandford1978}Blandford, R. D., Rees, M. J., 1978, in Wolfe A. M., ed., Pittsburgh ConferBence on BLLac Objects. Univ. Pittsburgh press, Pittsburgh, p. 328. \url{https://babel.hathitrust.org/cgi/pt?id=uc1.b4520824&view=1up&seq=328}

\bibitem[Blandford et al.(1990)]{Blandford1990}Blandford, R. D. 1990, in Active Galactic Nuclei, ed. T. J. L. Courvoisier $\&$ M. Mayor (Berlin:Springer), 161. \url{https://ui.adsabs.harvard.edu/abs/1990agn..conf..161B/abstract}

\bibitem[Blandford et al.(2019)]{Blandford2019}Blandford, R., Meier, D., Readhead, A., 2019, ARA$\&$ A, 57, 467. \url{https://doi.org/10.1146/annurev-astro-081817-051948}

\bibitem[Cao et al.(1999)]{Cao1999}Cao, X., Jiang, D. R. 1999, MNRAS, 307, 802. \url{https://doi.org/10.1046/j.1365-8711.1999.02657.x}

\bibitem[Cao et al.(2003)]{Cao2003}Cao, X. W. 2003, ApJ, 599, 147. \url{https://doi.org/10.1086/379240}

\bibitem[Calderone et al.(2013)]{Calderone2013}Calderone, G., Ghisellini, G., Colpi, M. $\&$ Dotti, M. 2013,MNRAS, 431, 210. \url{https://doi.org/10.1093/mnras/stt157}

\bibitem[Celotti et al.(1997)]{Celotti1997}Celotti, A., Padovani, P., Ghisellini, G., 1997, MNRAS, 286, 415. \url{https://doi.org/10.1093/mnras/286.2.415}

\bibitem[Chai et al.(2012)]{Chai2012}Chai, B., Cao, X. W., Gu, M. F, 2012, ApJ, 759, 114. \url{https://doi.org/10.1088/0004-637X/759/2/114}

\bibitem[Daly et al.(2018)]{Daly2018}Daly, R. A., Stout, D, A., Mysliwiec, J. N. 2018, ApJ, 863, 117. \url{https://doi.org/10.3847/1538-4357/aad08b}

\bibitem[Daly et al.(2019)]{Daly2019}Daly, R. A. 2019, ApJ, 886, 37. \url{https://doi.org/10.3847/1538-4357/ab35e6}

\bibitem[Daly et al.(2022)]{Daly2022}Daly, R. A. 2022, MNRAS, 517, 5144. \url{https://doi.org/10.1093/mnras/stac2976}

\bibitem[Falcke et al.(1994)]{Falcke1994}Falcke, H., Biermann, P. L., 1994, A$\&$A, 293, 665. \url{https://doi.org/10.48550/arXiv.astro-ph/9411096}

\bibitem[Falcke et al.(2004)]{Falcke2004}Falcke, H., K$\rm\ddot{o}$rding, E., Markoff, S. 2004, A$\&$A., 414, 895. \url{https://doi.org/10.1051/0004-6361:20031683}

\bibitem[Fan et al.(2018)]{Fan2018}Fan, X. L., Wu, Q. 2018, ApJ, 869, 133. \url{https://doi.org/10.3847/1538-4357/aaeece}

\bibitem[Fender et al.(2003)]{Fender2003}Fender, R. P., Gallo, E., Jonker, P. G. 2003, MNRAS, 343, L99. \url{https://doi.org/10.1046/j.1365-8711.2003.06950.x}

\bibitem[Fan et al.(2019)]{Fan2019}Fan, X. L., Wu, Q. 2019, ApJ, 879, 107. \url{https://doi.org/10.3847/1538-4357/ab25f1}

\bibitem[Fichtel et al.(1994)]{Fichtel1994}Fichtel, C. E., Bertsch, D. L., Chiang, J., et al. 1994, ApJS, 94, 551. \url{https://doi.org/10.1086/192082}

\bibitem[Francis et al.(1991)]{Francis1991}Francis, P. J., Hewett, P. C., Foltz, C. B., et al. 1991, ApJ, 373, 465. \url{https://doi.org/10.1086/170066}

\bibitem[Gallo et al.(2003)]{Gallo2003}Gallo, E., Fender, R. P., Pooley, G. G. 2003, MNRAS, 343, 60. \url{https://doi.org/10.1046/j.1365-8711.2003.06791.x}

\bibitem[Garofalo et al.(2010)]{Garofalo2010}Garofalo, D., Meier, D. L. 2010, MNRAS,406, 2047. \url{https://doi.org/10.1111/j.1365-2966.2010.16815.x}

\bibitem[Ghisellini et al.(2010)]{Ghisellini2010}Ghisellini, G., Tavecchio, F., Foschini, L., et al. 2010, MNRAS, 402, 497. \url{https://doi.org/10.1111/j.1365-2966.2009.15898.x}

\bibitem[Ghisellini et al.(2014)]{Ghisellini2014}Ghisellini, G., Tavecchio, F., Maraschi, L., Celotti, A., Sbarrato, T. 2014, Nature, 515, 376. \url{https://doi.org/10.1038/nature13856}

\bibitem[Godfrey et al.(2013)]{Godfrey2013}Godfrey, L. E. H., Shabala, S. S. 2013, ApJ, 767, 12. \url{https://doi.org/10.1088/0004-637X/767/1/12}

\bibitem[Gultekin et al.(2009)]{Gultekin2009}Gultekin, K., Cackett, E. M., Miller, J. M., et al. 2009, ApJ, 706, 404. \url{https://doi.org/10.1088/0004-637X/706/1/404}

\bibitem[Gultekin et al.(2019)]{Gultekin2019}Gultekin, K., King, A. L., Cackett, E. M., et al. APJ, 871, 80. \url{https://iopscience.iop.org/article/10.3847/1538-4357/aaf6b9}

\bibitem[Gu et al.(2009)]{Gu2009}Gu, M. F., Cao, X. W., Jiang, D. R., 2009, MNRAS, 396, 984. \url{https://doi.org/10.1111/j.1365-2966.2009.14758.x}

\bibitem[Heinz et al.(2003)]{Heinz2003}Heinz, S., Sunyaev, R, A. 2003, MNRAS, 343, L59. \url{https://doi.org/10.1046/j.1365-8711.2003.06918.x}

\bibitem[Ho et al.(2001)]{Ho2001}Ho, L. C., Peng, C. Y. 2001, APJ, 555, 650. \url{https://iopscience.iop.org/article/10.1086/321524}

\bibitem[King et al.(2013)]{King2013}King, A. L., Miller, J. M., Reynolds, M. T., et al. 2013, ApJL, 774, L25 \url{https://dx.doi.org/10.1038/nphys3724}

\bibitem[Kording et al.(2006a)]{Kording2006a}K$\rm\ddot{o}$rding, E., Falcke, H., Corbel, S. 2006, A$\&$A, 456, 439. \url{https://doi.org/10.1051/0004-6361:20054144}

\bibitem[Kording et al.(2006b)]{Kording2006b}K$\rm\ddot{o}$rding, E., Fender, R., Migliari, S. 2006, MNRAS, 369, 1451. \url{https://doi.org/10.1111/j.1365-2966.2006.10383.x}

\bibitem[Liu et al.(2006)]{Liu2006}Liu, Y., Jiang, D. R., Gu, M. F., 2006, ApJ, 637, 669. \url{https://doi.org/10.1086/498639}

\bibitem[Leon-Tavares et al.(2011)]{Leon2011a}Leon, T. J., Valtaoja, E., Chavushyan, V, H., et al. 2011, MNRAS, 411, 1127. \url{https://doi.org/10.1111/j.1365-2966.2010.17740.x}

\bibitem[Meier et al.(2001)]{Meier2001}Meier, D. L., 2001, ApJ, 548, L9. \url{https://doi.org/10.1086/318921}

\bibitem[Meier et al.(2002)]{Meier2002}Meier, D. L., 2002, New Astron. Rev, 46, 247. \url{https://doi.org/10.1016/S1387-6473(01)00189-0}

\bibitem[Merloni et al.(2003)]{Merloni2003}Merloni, A., Heinz, S., Matteo, T. 2003, MNRAS, 345, 1057. \url{https://doi.org/10.1046/j.1365-2966.2003.07017.x}

\bibitem[Miller et al.(2009)]{Miller2009}Miller, J. M., Reynolds, C. S., Fabian, A. C., Miniutti, G., Gallo, L. C. 2009, ApJ, 697, 900. \url{https://doi.org/10.1088/0004-637X/697/1/900}

\bibitem[Misner et al.(1973)]{Misner1973}Misner, C. W., Thorne, K. S., Wheeler, J. A. 1973, Gravitation (San Francisco: W.H.Freeman and Company), 3. \url{https://ui.adsabs.harvard.edu/abs/1973grav.book.....M/abstract}

\bibitem[Nemmen et al.(2012)]{Nemmen2012}Nemmen, R. S., Georganopoulos, M., Guiriec, S., et al. 2012, Science, 338, 1445. \url{https://doi.org/10.1126/science.1227416}

\bibitem[Netzer et al.(1990)]{Netzer1990}Netzer, H. 1990, in Active Galactic Nuclei (Berlin: Springer) ,57. \url{https://ui.adsabs.harvard.edu/abs/1990agn..conf...57N/abstract}

\bibitem[Nolan et al.(2012)]{Nolan2012}Nolan, P.L., Abdo, A. A., Ackermann, M., Ajello, M., et al. 2012, ApJS, 199, 31. \url{https://doi.org/10.1088/0067-0049/199/2/31}

\bibitem[Plotkin et al.(2012)]{Plotkin2012}Plotkin, R. M., Markoff, S., Kelly, B. C., Koerding, E., Anderson, S., F. 2012, MNRAS, 419, 267. \url{https://doi.org/10.1111/j.1365-2966.2011.19689.x}

\bibitem[Rawlings et al.(1991)]{Rawlings1991}Rawlings, S., Saunders, R., 1991, Nature, 349, 138. \url{https://doi.org/10.1038/349138a0}

\bibitem[Reynolds et al.(2014)]{Reynolds2014}Reynolds, C. S. 2014, SSRv, 183, 277. \url{https://doi.org/10.1007/s11214-013-0006-6}

\bibitem[Rees et al.(1984)]{Rees1984}Rees, M. J. 1984, ARA$\&$A, 22, 471. \url{https://doi.org/10.1146/annurev.aa.22.090184.002351}

\bibitem[Saikia et al.(2015)]{Saikia2015}Saikia, P., K$\rm\check{A}\rm\Grave{\Grave{u}}$rding, E., Falcke, H. 2015, MNRAS, 450, 2317. \url{https://doi.org/10.1093/mnras/stv731}

\bibitem[Saikia et al.(2016)]{Saikia2016}Saikia, P., K$\rm\check{A}\rm\Grave{\Grave{u}}$rding, E., Falcke, H. 2016, MNRAS, 461, 297. \url{https://doi.org/10.1093/mnras/stw1321}

\bibitem[Saikia et al.(2018)]{Saikia2018}Saikia, P., Kording, E., Coppejans, Deanne L., et al. 2018, A$\&$A, 616, 15 \url{https://doi.org/10.1051/0004-6361/201833233}

\bibitem[Sahakyan et al.(2021)]{Sahakyan2021}Sahakyan, N., Giommi, P. 2021, MNRAS, 502, 836 \url{https://doi.org/10.1093/mnras/stab011}

\bibitem[Sbarrato et al.(2012)]{Sbarrato2012}Sbarrato, T., Ghisellini, G., Maraschi, L., Colpi, M., 2012, MNRAS, 421, 1764. \url{https://doi.org/10.1111/j.1365-2966.2012.20442.x}

\bibitem[Scarpa et al.(1997)]{Scarpa1997}Scarpa, R., Falomo, R., 1997, A$\&$A, 325, 109. \url{https://articles.adsabs.harvard.edu/pdf/1997A\%26A...325..109S}

\bibitem[Shaw et al.(2012)]{Shaw2012}Shaw, M. S., Romani, R. W., Cotter, G., et al. 2012, ApJ, 748, 49. \url{https://doi.org/10.1088/0004-637X/748/1/49}

\bibitem[Shen et al.(2011)]{Shen2011}Shen, Y., Liu, X., Greene, J. E., et al. 2011, ApJS, 194, 45. \url{https://doi.org/10.1088/0067-0049/194/2/45}

\bibitem[Tavecchio et al.(2010)]{Tavecchio2010}Tavecchio, F., Ghisellini, G., Ghirlanda, G., Foschini, L., Maraschi, L., 2010, MNRAS, 401, 1570. \url{https://doi.org/10.1111/j.1365-2966.2009.15784.x}

\bibitem[Thorne et al.(1986)]{Thorne1986}Thorne, K. S., Price, R. H., Macdonald, D. A., Suen, W. M., Zhang, X. H., 1986, in Black Holes: The Membrane Paradigm(New Haven: Yale University Press), 67. \url{https://ui.adsabs.harvard.edu/abs/1986bhmp.book...67T/abstract}

\bibitem[Unal et al.(2015)]{Unal2020}$\rm\ddot{U}$nal, C ., Loeb, A. 2015, MNRAS, 495, 278. \url{https://doi.org/10.1093/mnras/staa1119}

\bibitem[Urry et al.(1995)]{Urry1995}Urry, C. M., Padovani, P., 1995, PASP, 107, 803. \url{https://doi.org/10.1086/133630}

\bibitem[Vestergaard et al.(2009)]{Vestergaard2009}Vestergaard, M., Osmer, P. S., 2009, ApJ, 699, 800. \url{https://doi.org/10.1088/0004-637X/699/1/800}

\bibitem[Vestergaard et al.(2006)]{Vestergaard2006}Vestergaard, M., Peterson, B. M., 2006, ApJ, 641, 689. \url{https://doi.org/10.1086/500572}

\bibitem[Wang et al.(2004)]{Wang2004}Wang, J. Min; Luo, B; Ho, L. C., 2014, ApJ, 797, 65. \url{https://doi.org/10.1088/0004-637X/797/1/65}

\bibitem[Wang et al.(2017)]{Wang2017}Wang, F. Y., $\&$Dai, Z. G., 2017, MNRAS, 470, 1101. \url{https://doi.org/10.1093/mnras/stx1292}

\bibitem[Wu et al.(2008)]{Wu2008}Wu, Q., Cao, X. 2008, ApJ, 678, 156. \url{https://doi.org/10.1086/591933 }

\bibitem[Woo et al.(2002)]{Woo2002}Woo, J. H., Urry, C. M., 2002, ApJ, 579, 530. \url{https://doi.org/10.1086/342878}

\bibitem[Willott et al.(1999)]{Willott1999}Willott, C. J., Rawlings, S., Blundell, K. M., et al. 1999, MNRAS, 309, 1017. \url{https://doi.org/10.1046/j.1365-8711.1999.02907.x}

\bibitem[Xie et al.(1991)]{Xie1991}Xie, G. Z., Liu, F. K., Liu, B. F.,et al. 1991, A$\&$A, 249, 65. \url{https://articles.adsabs.harvard.edu/pdf/1991A\%26A...249...65X}

\bibitem[Xie et al.(2004)]{Xie2004}Xie, G. Z., Zhou, S. B., Liang, E. W., 2004, AJ, 127, 53. \url{https://doi.org/10.1086/380218}

\bibitem[Xie et al.(2017)]{Xie2017}Xie, F. Z., Yuan, F., 2017, AJ, 836,104. \url{https://doi.org/10.3847/1538-4357/aa5b90}

\bibitem[Xiong et al.(2014)]{Xiong2014}Xiong, D. R., Zhang, X. 2014, MNRAS, 441, 3375. \url{https://doi.org/10.1093/mnras/stu755}

\bibitem[Xiong et al.(2015)]{Xiong2015}Xiong, D. R., Zhang, X., Bai, J. M., Zhang, H. J., 2015, MNRAS, 450, 3568. \url{https://doi.org/10.1093/mnras/stv812}

\bibitem[Zhang et al.(2017)]{Zhang2017}Zhang, X., Zhang, H. J., Zhang, X., Xiong, D. R. 2017, Ap$\&$SS, 362, 244. \url{https://doi.org/10.1007/s10509-017-3199-4}

\bibitem[Zhang et al.(2018)]{Zhang2018}Zhang, X., Zhang, H. J., Zhang, X. 2018, Ap$\&$SS, 363, 259. \url{https://doi.org/10.1007/s10509-018-3478-8}

\bibitem[Zhang et al.(2012)]{Zhang2012}Zhang, J., Liang, E. W., Zhang, S. N., Bai, J. M., 2012, ApJ, 752, 157. \url{https://doi.org/10.1088/0004-637X/752/2/157}

\bibitem[Zhang et al.(2023a)]{Zhang2023a}Zhang, X., Gao, Q. G., 2023, Ap$\&$SS, 368, 69. \url{https://doi.org/10.1007/s10509-023-04225-y}

\bibitem[Zhang et al.(2023b)]{Zhang2023b}Zhang, X., Gao, Q. G., 2023, RAA, 23, 125019. \url{https://doi.org/10.1088/1674-4527/ad020c}

\bibitem[Zhou et al.(2009)]{Zhou2009}Zhou, M., Cao, X., 2009, RAA, 9, 293. \url{https://doi.org/10.1088/1674-4527/9/3/003}



\end{thebibliography}
\end{document}